\documentclass[%
 reprint,
 amsmath,amssymb,
 aps,
]{revtex4-1}

\usepackage{graphicx}
\usepackage{dcolumn}
\usepackage{bm} 
\usepackage{amsmath}
\usepackage{hyperref}
\usepackage{physics} 
\usepackage[section]{placeins} 
\usepackage{xcolor}

\begin{document}

\preprint{APS/123-QED}

\title{Transport coefficients for the hot quark-gluon plasma at finite chemical potential $\mu_B$}

\author{Olga Soloveva}\email{soloveva@fias.uni-frankfurt.de}
\affiliation{Institute for Theoretical Physics, Johann Wolfgang
Goethe-Universit\"{a}t, Frankfurt am Main, Germany}

\author{Pierre Moreau}
\affiliation{Department of Physics, Duke University, Durham, North Carolina 27708, USA,\\
Institute for Theoretical Physics, Johann Wolfgang
Goethe-Universit\"{a}t, Frankfurt am Main, Germany}

\author{Elena Bratkovskaya}
\affiliation{GSI
Helmholtzzentrum f\"{u}r Schwerionenforschung GmbH,
 Darmstadt, Germany, \\ Institute for Theoretical Physics, Johann Wolfgang
Goethe-Universit\"{a}t, Frankfurt am Main, Germany}

\date{\today}

\begin{abstract}
We calculate transport coefficients of the quark-gluon plasma (QGP) within the  dynamical quasiparticle model (DQPM) by explicitly computing the  parton interaction rates as a function of temperature $T$ and baryon chemical potential $\mu_B$ on the basis of the DQPM couplings and partonic propagators. The latter are extracted from lattice QCD by matching the equation of state, entropy density and energy density at $\mu_B$= 0. For baryon chemical potentials  $0 \leq  \mu_B \leq 500 MeV$ we employ a scaling Ansatz for the effective coupling which was shown before to lead to thermodynamic consistent results in this range. We compute the ratio of the shear and bulk viscosities to the entropy density, i.e. $\eta/s$ and $\zeta/s$, the electric conductivity $\sigma_0/T$  as well as the baryon diffusion coefficient $\kappa_B$ and compare to related approaches from the literature. We find that the ratios $\eta/s$ and $\zeta/s$ as well as $\sigma_0/T$ are in accord with the results from lattice QCD at $\mu_B$=0 and only weakly depend on the ratio $T/T_c(\mu_B)$ where $T_c(\mu_B)$ denotes the critical temperature at finite baryon chemical potential.
\end{abstract}

\maketitle


\section{\label{sec:level1}Introduction}
The exploration of the QCD phase diagram at non-zero baryon density (or baryon chemical potential) is the mission driving a number of actual and future heavy-ion collision (HIC) experiments. Whereas at ultra-relativistic energies at the Large Hadron Collider (LHC) or the Relativistic Heavy-Ion Collider (RHIC) the quark-gluon plasma, which is created in the central interaction volume, has almost zero baryon chemical potential ($\mu_B \approx$ 0) and the transition from the deconfined state of quarks and gluons to the confined state of hadrons is a rapid but smooth crossover, the nature of the transition can not be understood from lattice QCD due to the fermionic sign problem at finite $\mu_B$. Only for small chemical potentials an expansion in terms of higher order susceptibilities gives some orientation, but at higher $\mu_B$ one presently has to employ theoretical models. In order to reach the region of the QCD phase diagram at finite baryon chemical potentials ongoing experiments are decreasing the center-of-mass collision energy and measuring observables at mid-rapidity. So far, the Beam Energy Scan (BES) program at RHIC has carried out a series of Au+Au collisions with the range of $\sqrt{s_{NN}}$ starting from $200$ GeV down to $7.7$ GeV. According to the statistical models the corresponding baryon chemical potential at chemical freeze-out is varying from $\mu_B \approx$ $20$ MeV at $\sqrt{s_{NN}}= 200$ GeV to $ \approx 420$ MeV at $\sqrt{s_{NN}}= 7.7$ GeV \cite{Adamczyk:BES17}. To study the existance of a critical end point(CEP) and possible effects of a first-order phase transition a rising of the baryon chemical potential can be achieved by further lowering the collision energy in the proposed BES phase III of fixed-target experiments and also in the future experiments at FAIR (Facility for Antiproton and Ion
Research) \cite{FAIR} and NICA (Nuclotron-based Ion Collider  fAcility) \cite{NICA} that particularly address the QGP phase diagram at moderate and higher $\mu_B$.

 Theoretical methods to explore QCD in Minkowski space for non-vanishing quark (or baryon) potentials are essentially effective approaches in which one can study the dominant properties of QCD in equilibrium, i.e. the thermodynamic quantities as well as transport coefficients. To this aim, the dynamical quasiparticle model (DQPM) has been introduced \cite{Cassing:2008nn} which is based on partonic propagators with sizeable imaginary parts of the self-energies incorporated. Whereas the real part of the self-energies can be attributed to a dynamically generated mass (squared) of the partons the imaginary parts contain the information about the interaction rates in the system. Furthermore, the imaginary parts of the propagators define the spectral functions of the degrees of freedom which might show characteristic quasiparticle peaks. A further advantage of a propagator based approach is that one can formulate a consistent thermodynamics \cite{Vanderheyden:1998,Blaizot:2000fc} as well as a causal theory for non-equilibrium configurations on the basis of Kadanoff-Baym (KB) equations. Transport coefficients are particularly interesting since they reveal the information about the interactions in the medium which in equilibrium can be characterized by a temperature $T$ and chemical potential $\mu_B$. There are also a lot of alternative effective models describing the partonic phase of the heavy-ion collision. While basically all of the effective models have similar equations of state (EoS), which match well with available lattice data, the transport coefficients can vary significantly \cite{Sasaki:2009,Bluhm:2011,Marty:NJL13}. Moreover, an exploration of transport coefficients of the hot and dense QGP provides useful information for hydrodynamical simulations of HICs.
 
 In this study we evaluate the $T$ and $\mu_B$ dependence of transport coefficients for the strongly interacting QGP on the basis of microscopic collision rates that are evaluated from the effective coupling and propagators of the DQPM on the tree level following our earlier work in Ref.~\cite{Pierre19}. A related approach to transport coefficients has been recently presented in Ref.~\cite{Mykhaylova:2019}.

 The paper is organized as follows: In Sec.\ref{sec2} we give a brief review of the basic parton properties in the DQPM and describe the calculation of the equation of state including a specification of the different contributions to the entropy density. Sec. \ref{sec3} is devoted to the actual computation of transport coefficients based on the relaxation time approximation, i.e. the shear and bulk viscosities, the electric conductivity as well as the baryon diffusion coefficient. We, furthermore, compare our results at $\mu_B=0$ to calculations from lattice QCD for $N_f=0$,  predictions from a Bayesian analysis, and estimates based on the Chapman-Enskog method. We close our study with conclusions in Sec. \ref{sec4}.

\section{\label{sec2}Parton properties in the DQPM}
In order to describe the partonic phase of heavy-ion collisions on the microscopic level, the dynamical quasiparticle model (DQPM) was introduced in Refs.~\cite{Cassing:2008nn,Review}. This effective model defines strongly-interacting quarks and gluons in terms of quasiparticles with single-particle (two-point) Green’s functions in the form

\begin{equation}
\label{propdqpm} G^{R} (\omega, {\bf p}) = \frac{1}{\omega^2 - {\bf
p}^2 - M^2 + 2 i \gamma \omega}
\end{equation}
using $\omega=p_0$ for energy.\\
The coupling (squared) $g^2=4 \pi \alpha_{s}$ regulates
the strength of the interaction  and enters the definition of the DQPM thermal masses and widths.
Here we follow  a procedure similar to Refs. \cite{Hamsa:PRC16,Hamsa:JModPhys16} to determine the
effective  coupling  (squared) $g^2$ as a function of temperature
$T$, i.e. the coupling is defined at $\mu_B = 0$ by a
parametrization of the entropy density from  lattice QCD in the
following way:

\begin{equation}
g^2(s/s_{SB}) = d \left( (s/s_{SB})^e -1 \right)^f,
\label{coupling_DQPM}
\end{equation}
with  the Stefan-Boltzmann entropy density $s_{SB}^{QCD} = 19/9\pi^2 T^3$ and the parameters d=169.934, e=-0.178434 and f=1.14631. In the following, we use a parametrization of the entropy density at $\mu_B = 0$ calculated by lQCD from Refs.~\cite{Borsanyi:2012cr,Borsanyi:2013bia} to determine the DQPM coupling $g^2$ as a function of temperature.

To obtain the coupling at finite baryon chemical potential $\mu_B$, the scaling hypothesis assumes that $g^2$ is a function of the ratio of the effective temperature $T^* = \sqrt{T^2+\mu^2_q/\pi^2}$ and the $\mu_B$-dependent critical temperature $T_c(\mu_B)$ as:

\begin{equation}
g^2(T/T_c,\mu_B) = g^2\left(\frac{T^*}{T_c(\mu_B)},\mu_B =0 \right),
\label{coupling}
\end{equation}
with $\mu_B=3\mu_q$, $T_c(\mu_B) = T_c \sqrt{1-\alpha \mu_B^2}$, where $T_c$ is
the critical temperature at vanishing chemical potential ($\approx 0.158$ GeV) and $\alpha = 0.974\ \text{GeV}^{-2}$ as in Ref. \cite{Pierre19}. It was shown in Ref. \cite{Thorsten} that this scaling Ansatz provides results for the partonic pressure $P$ and quark density $n_q$ that are practically equivalent to results from an integration of the Maxwell relations, that guarantee thermodynamic consistency for baryon chemical potentials less than 0.5 GeV. Thus, we limit our study to this range in $\mu_B$, which accordingly to the statistical model roughly correspond to the averaged $\mu_B$ probed in nucleus-nucleus collisions at $\sqrt{s_{NN}}\approx 5$ GeV \cite{Cleymans:2006}. We mention that at even lower bombarding energies (and higher $\mu_B$) the heavy-ion collisions dynamics was found to be dominated by hadronic degrees of freedom \cite{Palmese:2016} with a low sensitivity to the partonic phase in small space-time volumes.

With the coupling $g^2$ fixed from lQCD one can now specify the  masses of the dynamical quasiparticles, which are assumed to be given by the HTL  thermal masses in the asymptotic high-momentum regime, i.e. for gluons by

\begin{equation}
M^2_{g}(T,\mu_B)=\frac{g^2(T,\mu_B)}{6}\left(\left(N_{c}+\frac{1}{2}N_{f}\right)T^2
+\frac{N_c}{2}\sum_{q}\frac{\mu^{2}_{q}}{\pi^2}\right)\ ,
\label{Mg9}
\end{equation}
and for quarks (antiquarks) by

\begin{equation}
M^2_{q(\bar q)}(T,\mu_B)=\frac{N^{2}_{c}-1}{8N_{c}}g^2(T,\mu_B)\left(T^2+\frac{\mu^{2}_{q}}{\pi^2}\right)\ ,
\label{Mq9}
\end{equation}
where $N_{c}=3$ stands for the number of colors while $N_{f}\ (=3)$
denotes the number of (light) flavors. The dynamical masses (\ref{Mq9})
in the QGP are large compared to the bare masses of the light
($u,d$) quarks and adopted in the form (\ref{Mq9}) for the ($u,d$) quarks.
The strange quark has a larger bare mass which also enters to some
extent the dynamical mass $M_s(T)$. This essentially suppresses the channel $g \rightarrow s + {\bar s}$ relative to the channel $g \rightarrow u + {\bar u}$ or $d + {\bar d}$ and controls the strangeness ratio in the QGP. Empirically $M_s(T,\mu_B)= M_u(T,\mu_B)+ \Delta M = M_d(T,\mu_B)+ \Delta M$ where $\Delta M$ =
30 MeV has been used. This parameter has been fixed once in comparison to experimental data for the $K^+/\pi^+$ ratio in central Au+Au collisions at $\sqrt{s_{NN}}$ = 200 GeV. Furthermore, the quasiparticles in the DQPM have finite widths, which are adopted in the form \cite{Review}

\begin{equation}
\label{widthg}
\gamma_{g}(T,\mu_B) = \frac{1}{3}N_{c}\frac{g^2(T,\mu_B)T}{8\pi}\ln\left(\frac{2c}{g^2(T,\mu_B)}+1\right),
\end{equation}
\begin{equation}
\label{widthq}
\gamma_{q(\bar
	q)}(T,\mu_B)=\frac{1}{3}\frac{N^{2}_{c}-1}{2N_{c}}\frac{g^2(T,\mu_B)T}{8\pi}
\ln\left(\frac{2c}{g^2(T,\mu_B)}+1\right),
\end{equation}
where $c=14.4$  is related to a magnetic cut-off, which is an additional parameter of the DQPM. Furthermore, we assume that the width of the strange quark is the same as that for the light ($u,d$) quarks. At large temperatures the masses and widths grow linearly with the temperature whereas at temperatures close to $T_c(\mu_B)$ the DQPM masses are enhanced. Furthermore, the behavior at finite baryon chemical potential follows the one of the coupling $g^2(T,\mu_B)$ and thus a decrease is observed with increasing $\mu_B$. We mention that the finite width of the propagator $\gamma_i$ (\ref{widthg}),  (\ref{widthq}) is related to the collision rate of parton $i=q(\overline{q}),g$ by $\Gamma_i = 2 \gamma_i$. In this way we can check if the parametrizations  (\ref{widthg}),  (\ref{widthq}) are consistent with the collisional widths computed microscopically (see below).

With the quasiparticle properties (or propagators) fixed as described above, one can
evaluate the entropy density $s(T,\mu_B)$, the pressure $P(T,\mu_B)$ and energy
density $\epsilon(T,\mu_B)$ in a straight forward manner by starting with
the entropy density and number density in the propagator representation from Baym \cite{Vanderheyden:1998,Blaizot:2000fc},

\begin{gather}
s^{dqp} = \label{sdqp} \\
\begin{align*}
& - \int \frac{d\omega}{2 \pi} \frac{d^3p}{(2 \pi)^3} \left[ d_g\ \frac{\partial n_B}{\partial T} \left( \Im(\ln -\Delta^{-1})+ \Im \Pi \Re \Delta \right) \right. \\
& + \sum_{q=u,d,s} d_q\ \frac{\partial n_F(\omega-\mu_q)}{\partial T} \left( \Im(\ln -S_q^{-1})+ \Im \Sigma_q \Re S_q \right) \\
&  + \sum_{\bar{q}={\bar{u}},{\bar{d}},{\bar{s}}} \left. d_{\bar{q}}\ \frac{\partial n_{F}(\omega+\mu_q)}{\partial T} \left( \Im(\ln -S_{\bar{q}}^{-1})+ \Im \Sigma_{\bar{q}} \Re S_{\bar{q}} \right) \right]
\end{align*}
\end{gather}

\begin{gather}
 n^{dqp} = - \int \frac{d\omega}{2 \pi} \frac{d^3p}{(2 \pi)^3} \label{nbdqp} \\
\begin{align*}
&  \left[ \sum_{q=u,d,s} d_q\ \frac{\partial n_F(\omega-\mu_q)}{\partial \mu_q} \left( \Im(\ln -S_q^{-1})+ \Im \Sigma_q \Re S_q \right) \right. \\
& \left. +  \sum_{\bar{q}={\bar{u}},{\bar{d}},{\bar{s}}} d_{\bar{q}}\  \frac{\partial n_{F}(\omega+\mu_q)}{\partial \mu_q} \left( \Im(\ln -S_{\bar{q}}^{-1})+ \Im \Sigma_{\bar{q}} \Re S_{\bar{q}} \right)  \right] \ ,
\end{align*}
\end{gather}
~\\

where $n_B(\omega) = (\exp(\omega/T)-1)^{-1}$ and
$n_F(\omega-\mu_q) = (\exp((\omega-\mu_q)/T)+1)^{-1}$ denote the
Bose-Einstein and Fermi-Dirac distribution functions, respectively, while $\Delta
=(p^2-\Pi)^{-1}$, $S_q = (p^2-\Sigma_q)^{-1}$ and $S_{\bar q} =
(p^2-\Sigma_{\bar q})^{-1}$ stand for the full (scalar)
quasiparticle propagators of gluons $g$, quarks $q$ and antiquarks
${\bar q}$.  In Eq. (\ref{sdqp})-(\ref{nbdqp}) $\Pi$ and $\Sigma = \Sigma_q
\approx \Sigma_{\bar q}$ denote the (retarded) quasiparticle
self-energies. Furthermore, the number of transverse gluonic degrees-of-freedom is $d_g=2 \times (N_c^2-1)$
while for the fermion degrees-of-freedom we use $d_q= 2 \times N_c$ and $d_{\bar{q}}= 2 \times N_c$. \\
\begin{figure}[h!]
\center{\includegraphics[width=0.97\linewidth]{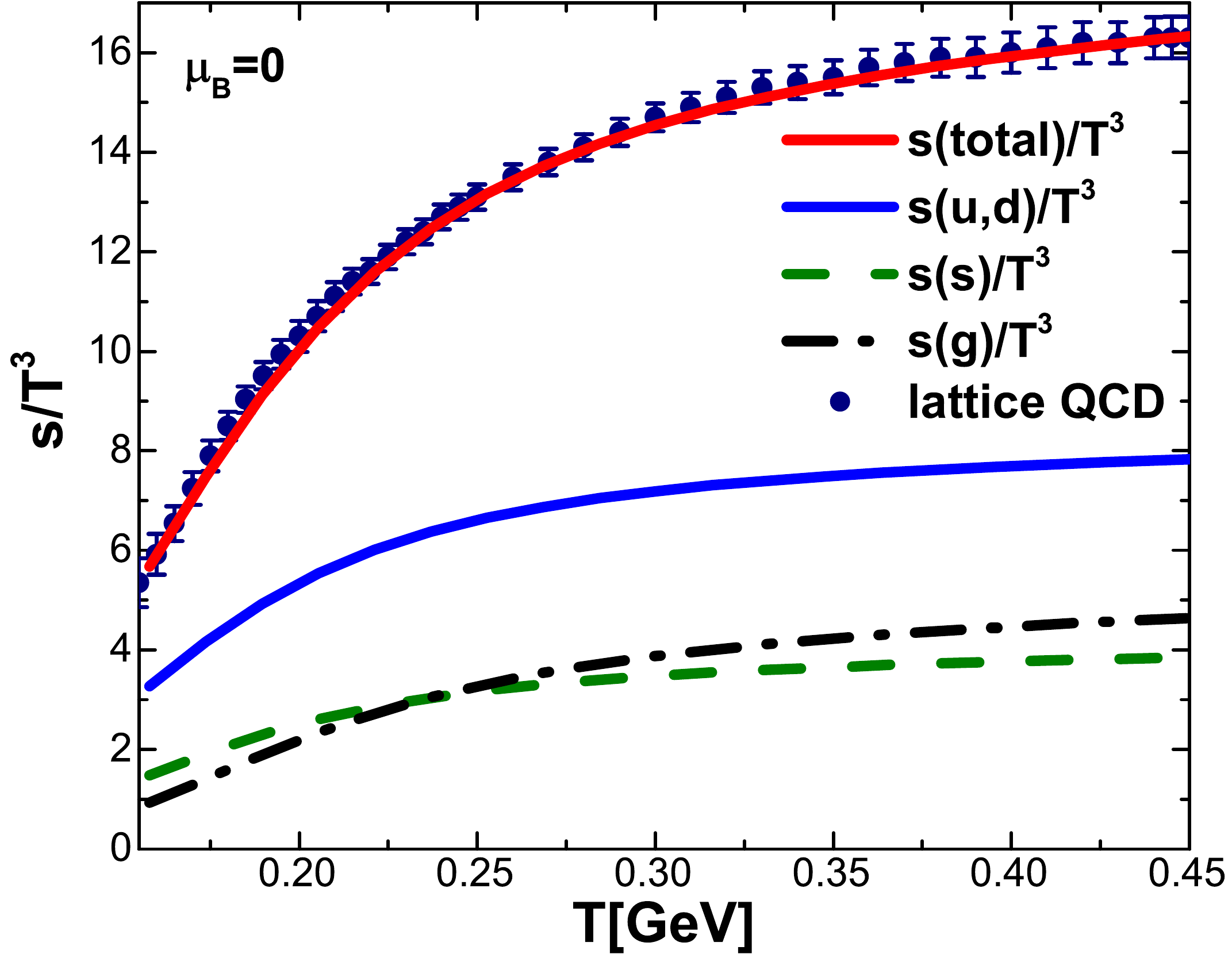} }
\caption{The scaled entropy density $s(T)/T^3$ from the DQPM for different quasiparticle species: light quarks (solid blue line), stange quarks (dot-dashed green line), gluons (dashed black line), total (red line) in comparison to the lQCD results from Ref. \cite{Borsanyi:2012cr} (full dots) for $\mu_B$ = 0. }
\label{fig:sdqp}
\end{figure}

In principle, $\Pi$ as well as $\Delta$ are Lorentz
tensors and should be evaluated in a nonperturbative framework. The
DQPM treats these degrees-of-freedom as independent scalar fields (for each color and spin projection)
with scalar self-energies  which are assumed to be identical for
quarks and antiquarks. This is expected to hold well for the entropy and number density.
Note that at finite quark chemical potential $\mu_q=\mu_B/3$ one has to deal with quarks and
antiquarks separately in Eqs. (\ref{sdqp})-(\ref{nbdqp}) since their abundance differs. \\

In case the real and imaginary parts of the propagators
$\Delta$ and $S$ are fixed, the entropy density
(\ref{sdqp}) and number density (\ref{nbdqp}) can be evaluated numerically. As we deal with a
grand-canonical ensemble, where the negative pressure is the thermodynamic potential, the Maxwell relations give

\begin{equation}
s =\frac{\partial P}{\partial T}\ \ ;\ \ n_B = \frac{\partial P}{\partial \mu_B} \ ,
\end{equation}
~\\
such that the pressure $P$ (and thus the thermodynamic potential) can be obtained by integration
of the entropy density $s$ over $T$ and of the baryon density $n_B$ over $\mu_B$ as:

\begin{gather}
P(T,\mu_B) = P(T_0,0) \label{pressure}\\
\begin{align*}
& \left. + \int_{T_{0}}^{T} s(T',0)\ dT' + \int_{0}^{\mu_B} n_B(T,\mu_B')\ d\mu_B' \right.
\end{align*}
\end{gather}

where one identifies the 'full' entropy density $s$ and baryon density $n_B$ with the quasiparticle entropy
density $s^{dqp}$ (\ref{sdqp}) and baryon density $n_B = n^{dqp}/3$ (\ref{nbdqp}). The starting point $T_{0}$ for the
integration in $T$ is chosen between 0.1 $< T < $ 0.15 GeV where
the entropy density is taken in accordance to the lattice QCD results from Ref. \cite{Borsanyi:2013bia}
in the hadronic sector. \\
The energy density $\epsilon$ then follows from the thermodynamical relation
\begin{equation}
\label{eps} \epsilon = T s - P +\mu_B n_B
\end{equation}
and thus is also fixed by the entropy $s(T,\mu_B)$ and baryon density $n_B(T,\mu_B)$ as well as the
interaction measure

\begin{equation}
\label{wint} I: = \epsilon - 3P = Ts - 4P + \mu_B n_B
\end{equation}
that vanishes for massless and noninteracting degrees of freedom at $\mu_B = 0$. \\

A good agreement between the resulting entropy density $s(T)$ (\ref{sdqp}), pressure $P(T)$ (\ref{pressure}),  energy density $\epsilon(T)$ (\ref{eps})  and interaction measure I(T) (\ref{wint}) from the DQPM and the lQCD results obtained by the BMW
group \cite{Borsanyi:2012cr,Borsanyi:2013bia} for $\mu_B = 0$  and $\mu_B$ = 400 MeV has been shown in Ref. \cite{Pierre19}.
For completeness and transparency the resulting entropy densities for different quasiparticles are shown in Fig. ~\ref{fig:sdqp}, along with the lattice QCD results from Ref. \cite{Borsanyi:2012cr}. The main contribution to the entropy density comes from the light quarks and antiquarks, while strange quarks have smaller contributions due to their larger mass. The contribution of gluons is of the same order, although these have an even larger mass and width, due to the degeneracy factor $d_g = 2(N_c^2-1) = 16$, which is larger than the strange quark degeneracy factor $d_s = 6$.

\section{\label{sec3}Transport coefficients}
In this section we calculate transport coefficients of the QGP in equilibrium within the DQPM thus extending our previous work in Ref. \cite{Pierre19}. We will consider the shear $\eta$ and bulk $\zeta$ viscosities, the electric conductivity $\sigma_0$ and the baryon diffusion coefficient $\kappa_B$. It has been found that in the quasi-particle approximation results for transport coefficients from the relaxation time approximation(RTA) of kinetic theory \cite{Hosoya:RTA,Chakraborty11,Kapusta,SGavin} and from the one-loop diagram calculations based on Kubo relations \cite{Kubo,Aarts:2002,Fraile:2006,Lang12} are very close. All transport coefficients, which we will describe later on, have been calculated within the relaxation time approximation(RTA). The first step in the calculation of transport coefficients within the RTA framework is the evaluation of relaxation times, which are supposed to depend on momenta, temperature, and baryon chemical potential. We have considered two cases for the relaxation times for quarks and gluons:
\begin{align}
& \left. 1)\tau_i(\mathbf{p},T,\mu_B) = \frac{1}{\Gamma_i(\mathbf{p},T,\mu_B)} \right.  \nonumber \\
& \left. 2)\tau_i(T,\mu_B) = \frac{1}{2\gamma_i(T,\mu_B)},\right.  
\end{align}
where $\Gamma_i(\mathbf{p},T,\mu_B)$ is the parton interaction rate based on the microscopic differential cross sections computed in Ref. \cite{Pierre19}, while $\gamma_i(T,\mu_B)$ are the parton spectral widths from (\ref{widthg}), (\ref{widthq}). For a detailed description of the differential partonic cross sections at finite $T$ and $\mu_B$ on the level of tree diagrams we refer the reader to Ref. \cite{Pierre19}. We briefly recall that
in the on-shell case (energies of the particles are taken to be $E^2 = \mathbf{p}^2 + M^2$ where $M$ is the pole mass) the collisional widths are calculated as follows:

\begin{align}
\Gamma^{\text{on}}_i & (\mathbf{p}_i, T,\mu_q) = \frac{1}{2E_i} \sum_{j=q,\bar{q},g} \int \frac{d^3p_j}{(2\pi)^3 2E_j}\ d_j\ f_j(E_j,T,\mu_q)  \nonumber \\
& \ \ \ \ \ \ \ \ \ \ \ \ \times  \int \frac{d^3p_3}{(2\pi)^3 2E_3}  \int \frac{d^3p_4}{(2\pi)^3 2E_4} (1\pm f_3) (1\pm f_4) \nonumber \\
& \ \ \ \ \ \ \ \times |\bar{\mathcal{M}}|^2 (p_i,p_j,p_3,p_4)\ (2\pi)^4 \delta^{(4)}\left(p_i + p_j -p_3 -p_4 \right) \nonumber \\
= & \sum_{j=q,\bar{q},g} \int \frac{d^3p_j}{(2\pi)^3}\ d_j\ f_j\ v_{\text{rel}} \int d\sigma^{\text{on}}_{ij \rightarrow 34}\ (1\pm f_3) (1\pm f_4) ,
\label{Gamma_on}
\end{align}
where $d_j$ is the degeneracy factor for spin and color (for quarks $d_q = 2 \times N_c$ and for gluons $d_g =2 \times (N_c^2-1)$), and with the shorthand notation $f_j = f_j(E_j,T,\mu_q)$ for the distribution functions. In Eq. (\ref{Gamma_on}) and in all this section, the notation $\sum_{j=q,\bar{q},g}$ includes the contribution from all possible partons which in our case are the gluons and the (anti-)quarks of three different flavors ($u,d,s$). Furthermore, $v_{\text{rel}}$ denotes the relative velocity of the colliding partons whereas $\sigma^{\text{on}}_{ij \rightarrow 34}$ stand for the differential cross sections computed in Ref. \cite{Pierre19}.

It is interesting to evaluate the parton relaxation times as a function of temperature $T$ and chemical potential $\mu_B$ times. To this aim we calculate the average width of the partons $i$, we finally have to average its interaction rate (\ref{Gamma_on}) over its momentum distribution,
\begin{align}
\Gamma^{\text{on}}_i(T,\mu_q) & = \frac{d_i}{n_i^{\text{on}}(T,\mu_q)} \int \frac{d^3p_i}{(2\pi)^3}\ f_i(E_i,T,\mu_q)  \nonumber \\ 
& \times \Gamma^{\text{on}}_i(\mathbf{p}_i,T,\mu_q)
\label{Gamma_on_avT}
\end{align}
~\\
with the on-shell density of partons $i$ at  $T$ and $\mu_q$ given by
\begin{equation}
n_i^{\text{on}}(T,\mu_q) = d_i \int \frac{d^3p_i}{(2\pi)^3}\  f_i(E_i,T,\mu_q) .
\label{n_on}
\end{equation}
~\\
In fact, as seen from Eq. (\ref{Gamma_on}), the interaction rate of particle $i$ is directly proportional to the density of the colliding partner $j$ and its degeneracy factor $d_j$, and to their interaction cross section $\sigma_{ij}$ as:

\begin{equation}
\Gamma_i \propto \sum_{j} d_j\ f_j\ \sigma_{ij} .
\end{equation}
If we consider e.g.  $u-$quark scattering, we obtain for all the possible interaction channels:
\begin{itemize}
	\item {(1)} : $ uu \rightarrow uu ;\ \Gamma_{uu \rightarrow uu} \propto d_u\ f_u\ \sigma_{uu \rightarrow uu}$ (t + u channels)
	\item {(2)} : $ u\bar{u} \rightarrow u\bar{u} ;\ \Gamma_{u\bar{u} \rightarrow u\bar{u}} \propto d_{\bar{u}}\ f_{\bar{u}}\ \sigma_{u\bar{u} \rightarrow u\bar{u}}$ (t + s channels)
	\item {(3)} : $ u\bar{u} \rightarrow d\bar{d} ;\ \Gamma_{u\bar{u} \rightarrow d\bar{d}} \propto d_{\bar{u}}\ f_{\bar{u}}\ \sigma_{u\bar{u} \rightarrow d\bar{d}}$ (s channel)
	\item {(4)} : $ u\bar{u} \rightarrow s\bar{s} ;\ \Gamma_{u\bar{u} \rightarrow s\bar{s}} \propto d_{\bar{u}}\ f_{\bar{u}}\ \sigma_{u\bar{u} \rightarrow s\bar{s}}$ (s channel)
	\item {(5)} : $ ud \rightarrow ud ;\ \Gamma_{ud \rightarrow ud} \propto d_d\ f_d\ \sigma_{ud \rightarrow ud}$ (t channel)
	\item {(6)} : $ u\bar{d} \rightarrow u\bar{d} ;\ \Gamma_{u\bar{d} \rightarrow u\bar{d}} \propto d_{\bar{d}}\ f_{\bar{d}}\ \sigma_{u\bar{d} \rightarrow u\bar{d}}$ (t channel)
	\item {(7)} : $ us \rightarrow us ;\ \Gamma_{us \rightarrow us} \propto d_s\ f_s\ \sigma_{us \rightarrow us}$ (t channel)
	\item {(8)} : $ u\bar{s} \rightarrow u\bar{s} ;\ \Gamma_{u\bar{s} \rightarrow u\bar{s}} \propto d_{\bar{s}}\ f_{\bar{s}}\ \sigma_{u\bar{s} \rightarrow u\bar{s}}$ (t channel)
	\item {(9)} : $ ug \rightarrow ug ;\ \Gamma_{ug \rightarrow ug} \propto d_g\ f_g\ \sigma_{ug \rightarrow ug}$ (t channel).
\end{itemize}
~\\
Adding up all the contributions, we get for the light quark total interaction rate:

\begin{align}
\Gamma_u & = \sum_{q} { \Gamma_{uq} } + \sum_{\bar{q}} { \Gamma_{u\bar{q}}} + { \Gamma_{ug}} \label{gamu} \\
& = (1) + (5) +  (7) +  (2) +  (3) +  (4) + (6) +  (8) + (9). \nonumber
\end{align}
~\\
Similarly for a gluon $g$, the possible interaction channels are:
\begin{itemize}
	\item {(10)} : $ gu \rightarrow gu ;\ \Gamma_{gu \rightarrow gu} \propto d_u\ f_u\ \sigma_{gu \rightarrow gu}$ \ \ \   = \ \ \  $ gd \rightarrow gd ;\ \Gamma_{gd \rightarrow gd} \propto d_d\ f_d\ \sigma_{gd \rightarrow gd}$ (t + u + s channels)
	\item {(11)} : $ g\bar{u} \rightarrow g\bar{u} ;\ \Gamma_{g\bar{u} \rightarrow g\bar{u}} \propto d_{\bar{u}}\ f_{\bar{u}}\ \sigma_{g\bar{u} \rightarrow g\bar{u}}$ \ \ \   = \ \ \  $ g\bar{d} \rightarrow g\bar{d} ;\ \Gamma_{g\bar{d} \rightarrow g\bar{d}} \propto d_{\bar{d}}\ f_{\bar{d}}\ \sigma_{g\bar{d} \rightarrow g\bar{d}}$  (t + u + s channels)
	\item {(12)} : $ gs \rightarrow gs ;\ \Gamma_{gs \rightarrow gs} \propto d_s\ f_s\ \sigma_{gs \rightarrow gs}$ (t + u + s channels)
	\item {(13)} : $ g\bar{s} \rightarrow g\bar{s} ;\ \Gamma_{g\bar{s} \rightarrow g\bar{s}} \propto d_{\bar{s}}\ f_{\bar{s}}\ \sigma_{g\bar{s} \rightarrow g\bar{s}}$ (t + u + s channels)
	\item {(14)} : $ gg \rightarrow gg ;\ \Gamma_{gg \rightarrow gg} \propto d_g\ f_g\ \sigma_{gg \rightarrow gg}$ (t + u + s channels + 4 point amplitude).
\end{itemize}
~\\
Adding up all the contributions, we get for the gluon total interaction rate:

\begin{align}
\Gamma_g & = \sum_{q} { \Gamma_{gq} } + \sum_{\bar{q}} {\Gamma_{g\bar{q}}} + { \Gamma_{gg}} \label{gamg} \\
& = 2\times (10) +  (12) + 2\times (11) + (13) +  (14). \nonumber
\end{align}
\begin{figure}[h!]
\begin{minipage}[h]{0.97\linewidth}
\center{\includegraphics[width=1\linewidth]{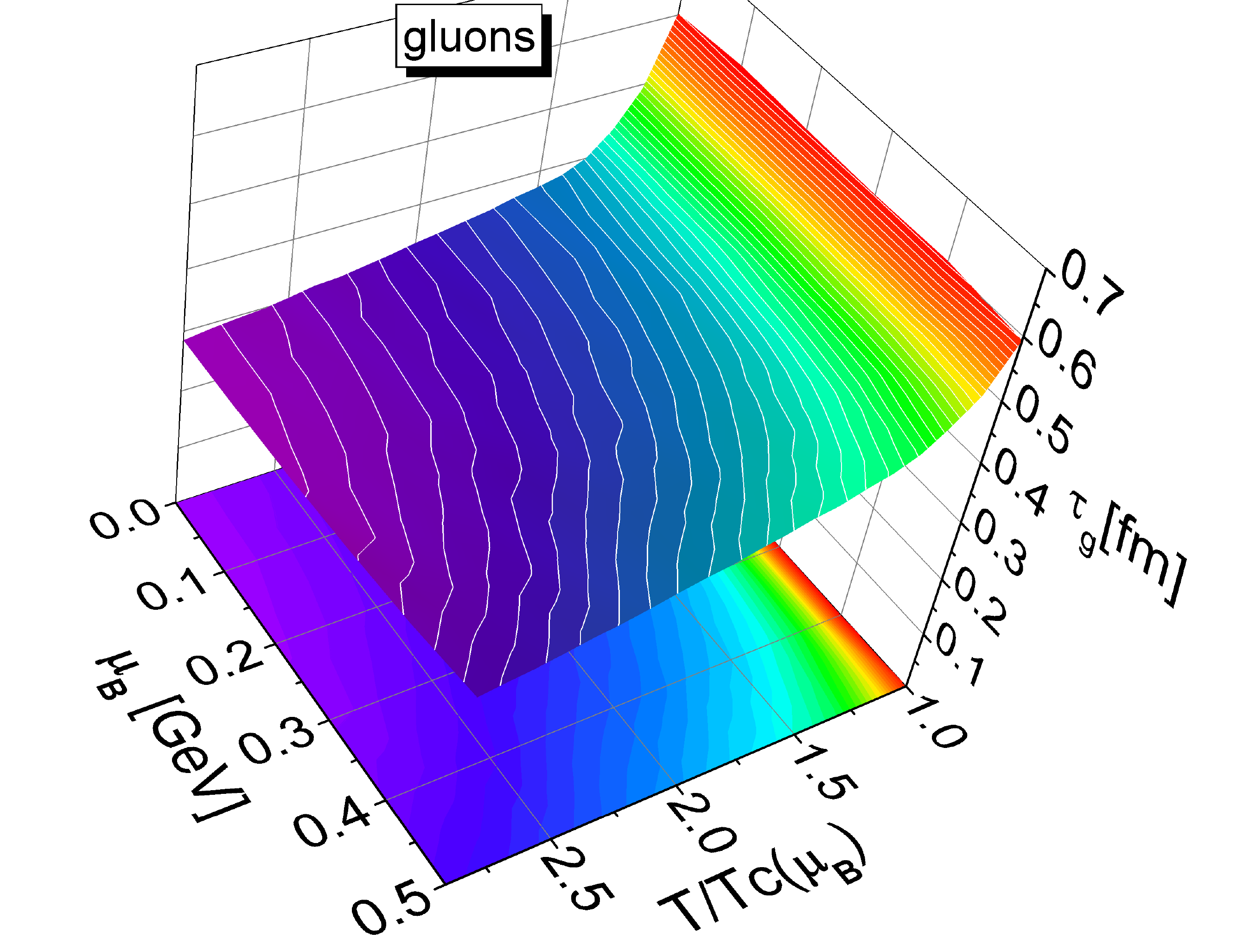} \\ a) }
\end{minipage}
\hfill
\begin{minipage}[h]{0.97\linewidth}
\center{\includegraphics[width=1\linewidth]{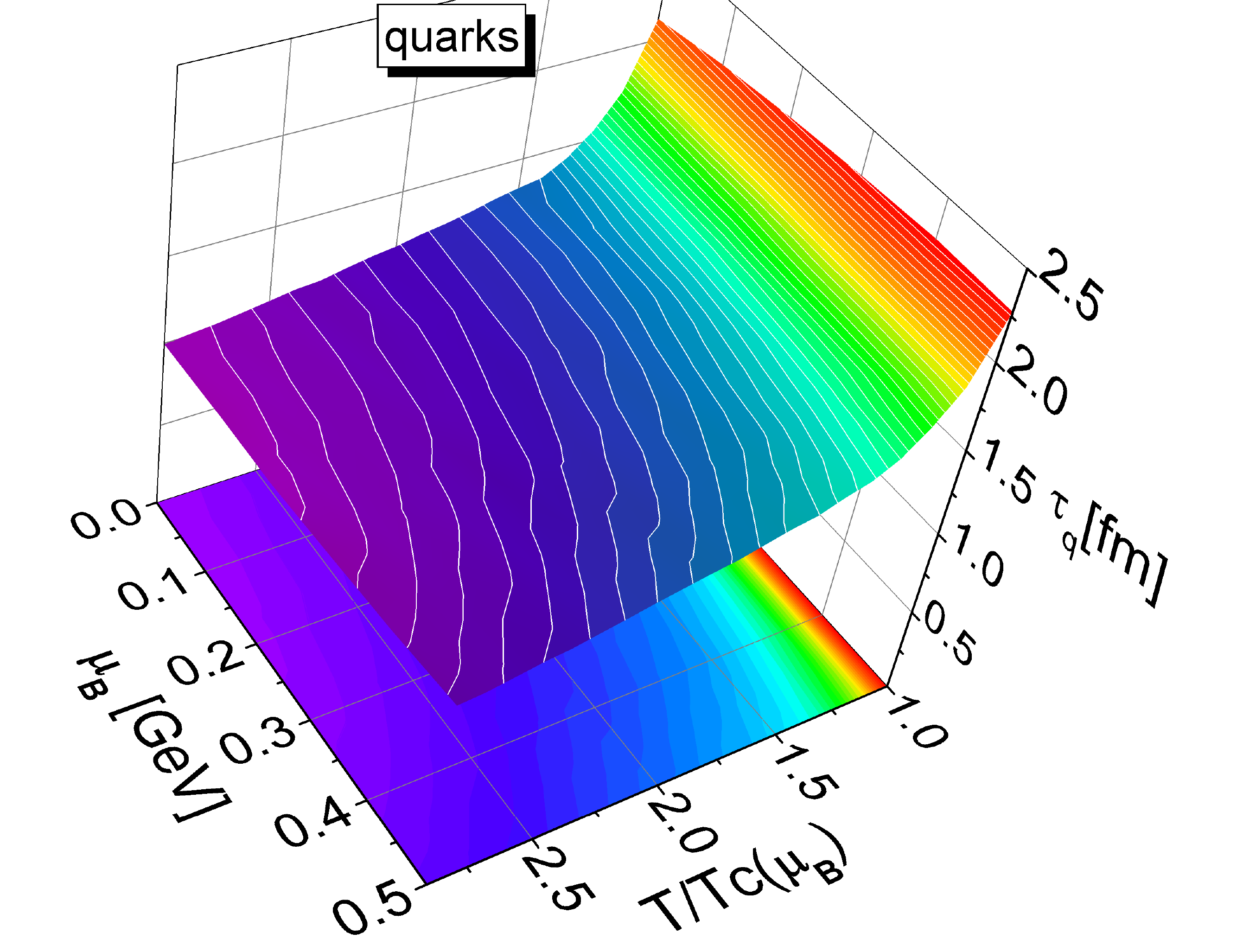} \\ b) }
\end{minipage}
\caption{Relaxation time of a gluon (a) and quark (b) as a function of the scaled temperature $T/T_c(\mu_B)$ and the baryon chemical potential $\mu_B$ evaluated by the average parton interaction rate from Eq. (\ref{Gamma_on_avT}) }
\label{fig:trelax}
\end{figure}
In Ref.~\cite{Pierre19} we have shown explicitely the quark and gluon off-shell interaction rates as a function of the scaled temperature $T/T_c(\mu_B)$ and chemical potential $\mu_B$.
While the dependencies on temperature are similar for fixed $\mu_B$ we see a general slight decrease in the total widths with $\mu_B$ for fixed temperature; the parton relaxation times - evaluated by $\Gamma_i^{-1}$ - then slightly increase with $\mu_B$. Fig.~\ref{fig:trelax} gives an overview of the relaxation time of a gluon (a) and quark (b) as a function of the scaled temperature $T/T_c(\mu_B)$ and chemical potential $\mu_B$. The gluon relaxation time is about $0.3-0.4$ fm/c in the region $1.5 T_c \leq T\leq 3 T_c$, which is significantly smaller than the quark relaxation time, which is about $1.0-1.5$ fm/c. Since the transport coefficients are directly proportional to the relaxation times, it is clear that the main contribution to the transport coefficients in the RTA stems from quarks and antiquarks.

\subsection{Shear viscosity}
\begin{figure}[h!]
\center{\includegraphics[width=0.85\linewidth]{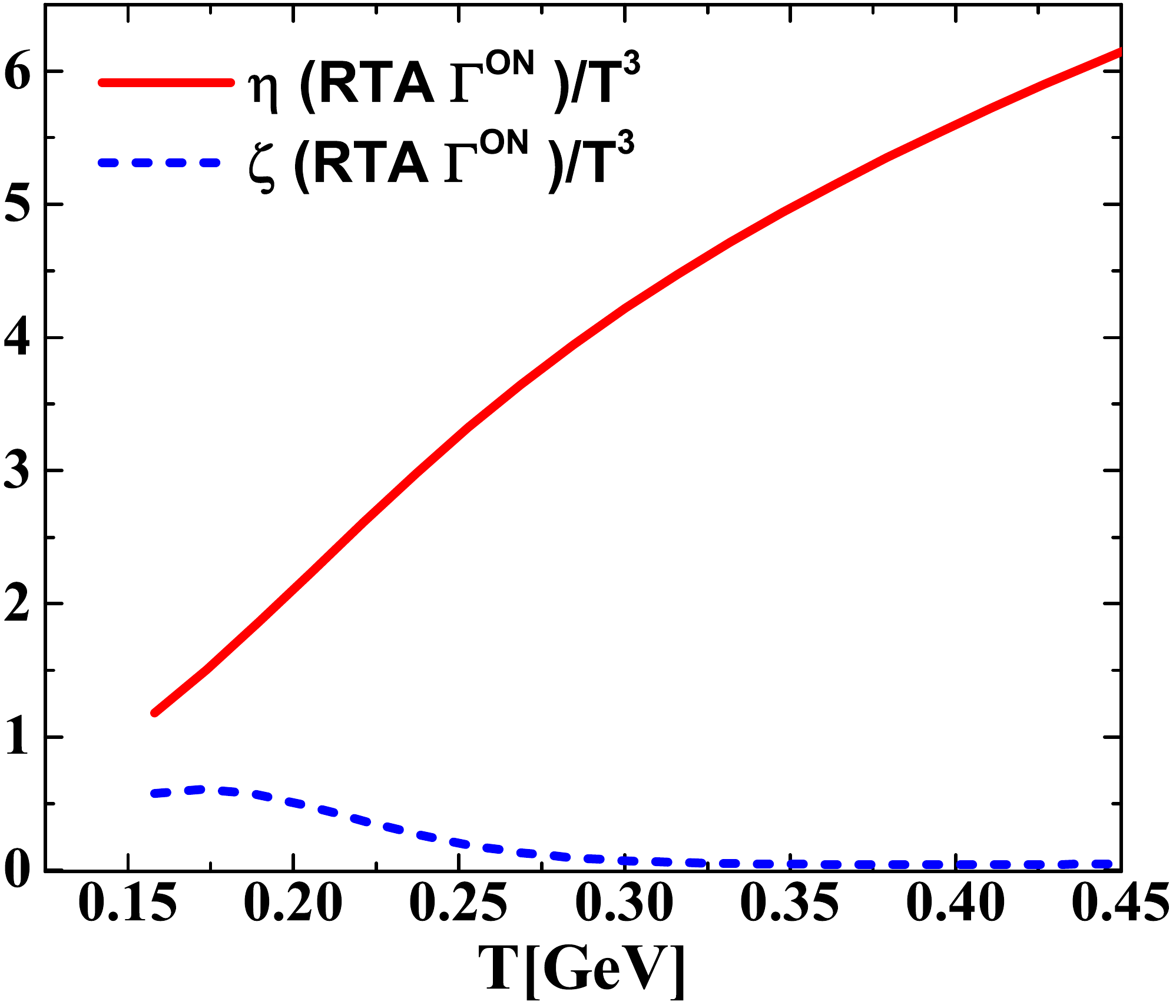} }
\caption{Ratio of shear $\eta/T^3$ and bulk $\zeta/T^3$ viscosities to the temperature  cubed as a function of temperature for $\mu_B$ = 0.  The solid red line and the dashed blue line show the DQPM results for the shear and bulk ratos accordingly, using the parton  interaction rate $\Gamma_i(\mathbf{p},T,\mu)$  for the relaxation time. }
\label{fig:viscT3}
\end{figure}

The shear viscosity to entropy density ratio $\eta/s$ for the QGP - created in the central region of high-energy heavy-ion collisions at Super ProtonSynchrotron (SPS) and Relativistic Heavy-Ion Collider (RHIC) - was predicted to be very low \cite{Shuryak,Gyulassy,Heinz}. While the ratio $\eta/s$ is expected to be small the shear viscosity $\eta$  as well as the entropy $s$ of partonic system are high and scale with the temperature as $\propto T^3$ as shown in Fig.~\ref{fig:sdqp} and  Fig.~\ref{fig:viscT3}. 

One way to evaluate the viscosity coefficients of partonic matter is the Kubo formalism \cite{Kubo,Aarts:2002,Fraile:2006,Lang12}, which was used to calculate the viscosities for a previous version of the DQPM within the PHSD transport approach in a box with periodic boundary conditions (cf. Ref. \cite{Ozvenchuk13:kubo}). We here focus on the calculation of the shear viscosity based on the RTA \cite{Kapusta} which reads:

\begin{align}
\eta^{\text{RTA}}(T,\mu_B)  = \frac{1}{15T} \sum_{i=q,\bar{q},g} \int \frac{d^3p}{(2\pi)^3} \frac{\mathbf{p}^4}{E_i^2}    \label{eta_on}  \tau_i(\mathbf{p},T,\mu_B) \\
\times  d_i  (1 \pm f_i) f_i , \nonumber
\end{align}
where  $d_q = 2N_c = 6$ and $d_g = 2(N_c^2-1) = 16$ are degeneracy factors for spin and color in case of quarks and  gluons , $\tau_i$ are the relaxation times. In extension to our previous studies in Refs. \cite{Hamsa:PRC16,Hamsa:JModPhys16,Marty:NJL13} we here include the Pauli-blocking and Bose enhancement factors, respectively.

\begin{figure}[h!]
\begin{minipage}[h]{0.85\linewidth}
\center{\includegraphics[width=0.97\linewidth]{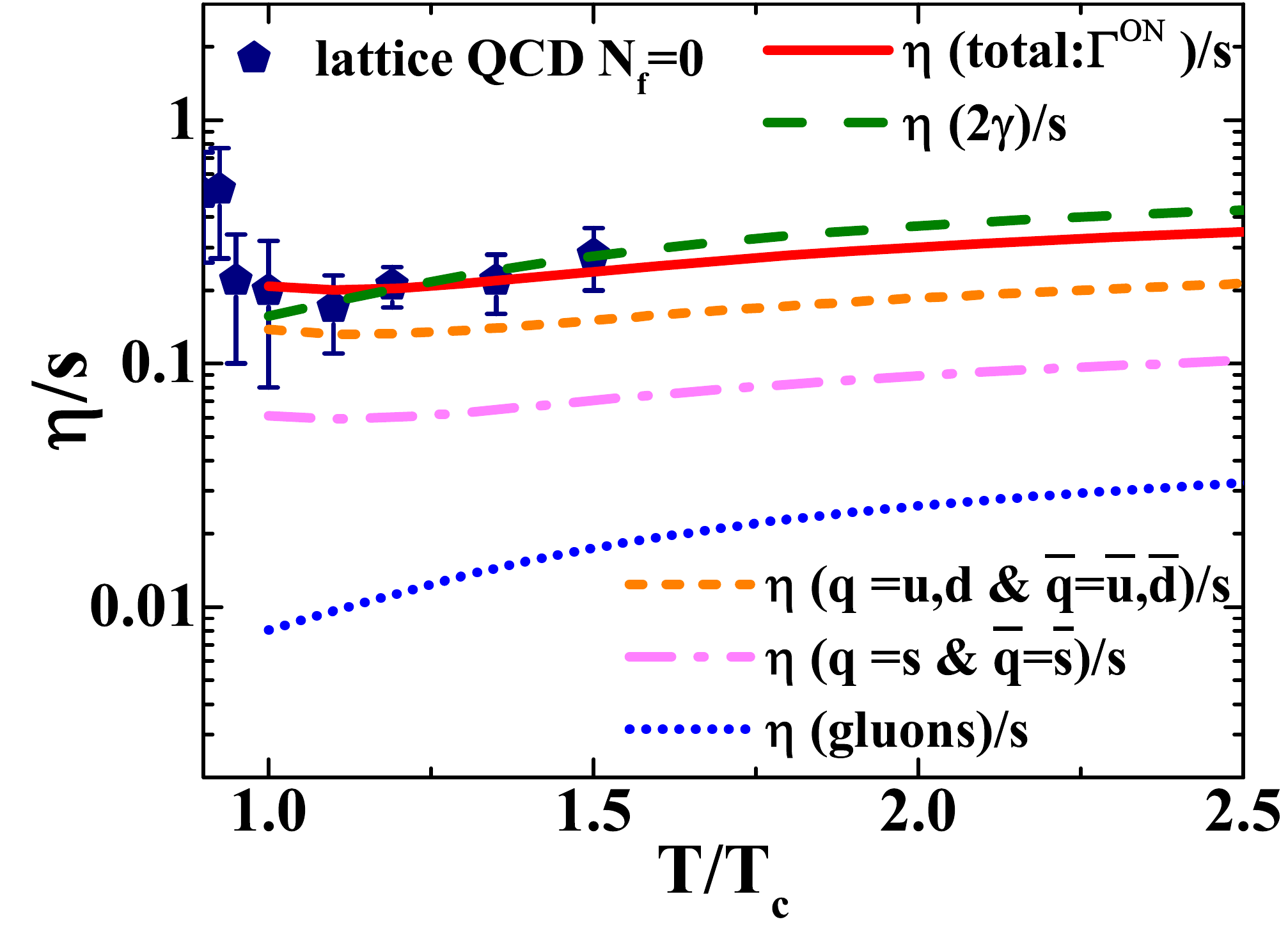} \\ a) $\mu_B$ =0 }
\end{minipage}
\hfill
\begin{minipage}[h]{0.9\linewidth}
\center{\includegraphics[width=0.9\linewidth]{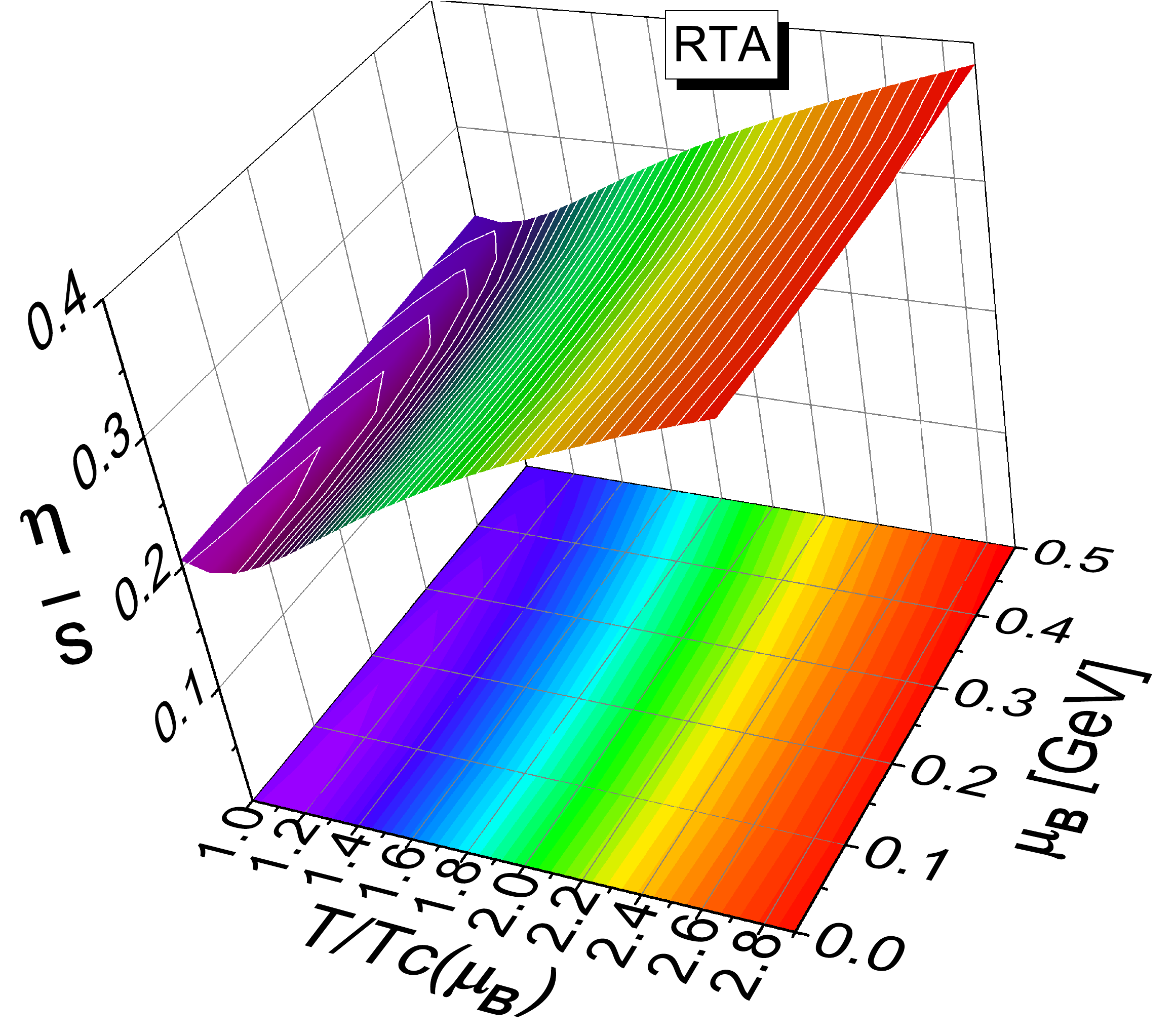} \\ b)  $\mu_B$ dependence}
\end{minipage}
\caption{Ratio of shear viscosity to entropy density a) as a function of scaled temperature $T/T_c$ for $\mu_B=0$ and b) for  non-zero $\mu_B$ as a function of the scaled temperature $T/T_c(\mu_B)$ and the baryon chemical potential $\mu_B$. The lines shows the DQPM result from Eq. (\ref{eta_on}) using the interaction rate $\Gamma_i(\mathbf{p},T,\mu)$ for different quasiparticle species: light quarks and anti-quarks (short dashed orange line), strange quarks and anti-quarks (dot-dashed magenta line), gluons (dotted blue line). The solid red line and the dashed green line show the DQPM results for total ratios of viscosity to entropy density using the parton  interaction rate $\Gamma_i(\mathbf{p},T,\mu)$  and the spectral width $2\gamma_i(T,\mu)$ for the relaxation time. The pentagons show the lQCD data for pure SU(3) gauge theory taken from Ref.~\cite{Astrakhantsev:2017}. }
\label{fig:etas}
\end{figure}

The ratio of the shear viscosity to entropy density $\eta/s$ has been presented before in Ref.~\cite{Pierre19}; it increases with an increase of the scaled temperature. The actual values for the ratio $\eta/s$  are in a good agreement with the gluodynamic lattice QCD calculations at $\mu_B=0$ from Ref.~\cite{Astrakhantsev:2017}. We find that the ratio $\eta/s$ does not vary  much with $\mu_B$ and has a similar behavior as a function of temperature for all $\mu_B$ considered.
The approximation (\ref{eta_on}) of the shear viscosity is found to be very close to the one from the Kubo formalism  \cite{Pierre19} indicating that the quasiparticle limit ($\gamma \ll M$) holds in the DQPM.

The ratio of the shear viscosity to entropy density at $\mu_B=0$ is shown in Fig.~\ref{fig:etas} (a) in comparison to the lattice QCD calculation for $N_f=0$ from Ref.~\cite{Astrakhantsev:2017}. The ratio $\eta/s$ at $\mu_B=0$ also is in a good agreement with the predictions from a Bayesian analysis of experimental heavy-ion data from Ref.~\cite{Sbass:2017}. In extension to Ref. \cite{Pierre19} we also show the separate contributions to $\eta/s$  for various quasiparticle species to the total ratio: light quarks and anti-quarks (short dashed orange line), strange quarks and anti-quarks (dot-dashed magenta line), gluons (dotted blue line). The solid red line correspond to the ratio of the total shear viscosity to entropy density. Smaller values of the shear viscosity are observed for the gluons than for the quarks as expected since the gluon relaxation time is approximately twice smaller than the quark relaxation time, and the masses of the gluons are approximately twice higher than quarks masses, which effects the shear viscosity via the factor $1/E_i^2$. The light quarks and anti-quarks give the main contribution to the total ratio $\sim 60$ \%. The strange quarks and anti-quarks contribute by $\sim 30$ \%, while the gluon contribution is about $\sim 10$ \%. The differences for the shear viscosity of different quark flavour is essentially due to the mass difference. We mention that the hierarchy obtained here is in a good agreement with the recent calculations for the shear viscosity at $\mu_B=0$ in the quasiparticle model of Ref.~\cite{Mykhaylova:2019}. It is worth to note that in Ref.~\cite{Mykhaylova:2019} only the on-shell case is considered where quasiparticles don't have widths and the couplings are higher than in the DQPM, which leads to an increase of the relaxation times and the shear viscosity to entropy density ratio $\eta/s$.

\subsection{Bulk viscosity}
The hydrodynamical simulations of ultrarelativistic heavy-ion collisions predicted that the bulk viscosity of the QGP should be non-zero, at least in the vicinity of the phase transition \cite{Ryu}.
In this study we evaluate the bulk viscosity of the partonic phase within the RTA following Ref.~\cite{Kapusta}:
\begin{align}
\zeta^{\text{RTA}}(T,\mu_B)= \frac{1}{9T} \sum_{i=q,\bar{q},g}\int \frac{d^3p}{(2\pi)^3}
  \label{zeta_on} \tau_i(\mathbf{p},T,\mu_B) \\
\times  \frac{ d_i  (1 \pm f_i) f_i  }{E_i^2}\left(\mathbf{p}^2-3c_s^2\left(E_i^2-T^2\frac{dm_i^2}{dT^2}\right)\right)^2 , \nonumber
\end{align}
where $c_s^2$ is the speed of sound squared, $\frac{dm_i^2}{dT^2}$ is the DQPM parton mass derivative which becomes large close to the critical temperature $T_c$.
The DQPM results for the viscosities over temperature cubed are showed at Fig.~\ref{fig:viscT3}, where  the parton  interaction rate $\Gamma_i(\mathbf{p},T,\mu_B)$ was used for the relaxation time. The solid red line corresponds to the ratio of the shear viscosity to the temperature cubed $\eta/T^3$, while the dashed blue line shows the ratio of bulk viscosity to the temperature cubed $\zeta/T^3$.

\begin{figure}[h!]
\begin{minipage}[h]{0.85\linewidth}
\center{\includegraphics[width=0.97\linewidth]{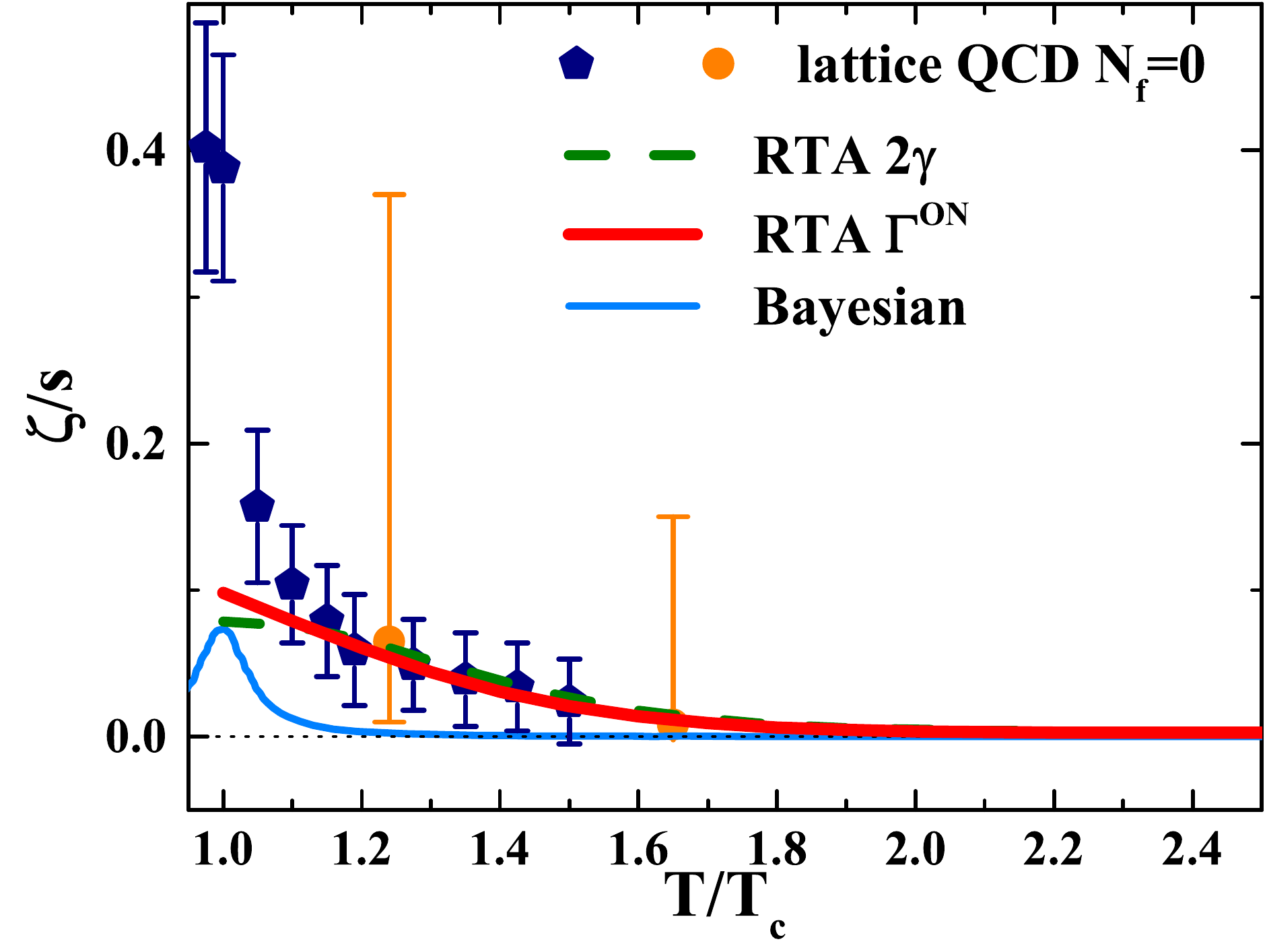} \\ a) $\mu_B$ =0 }
\end{minipage}
\hfill
\begin{minipage}[h]{0.9\linewidth}
\center{\includegraphics[width=0.95\linewidth]{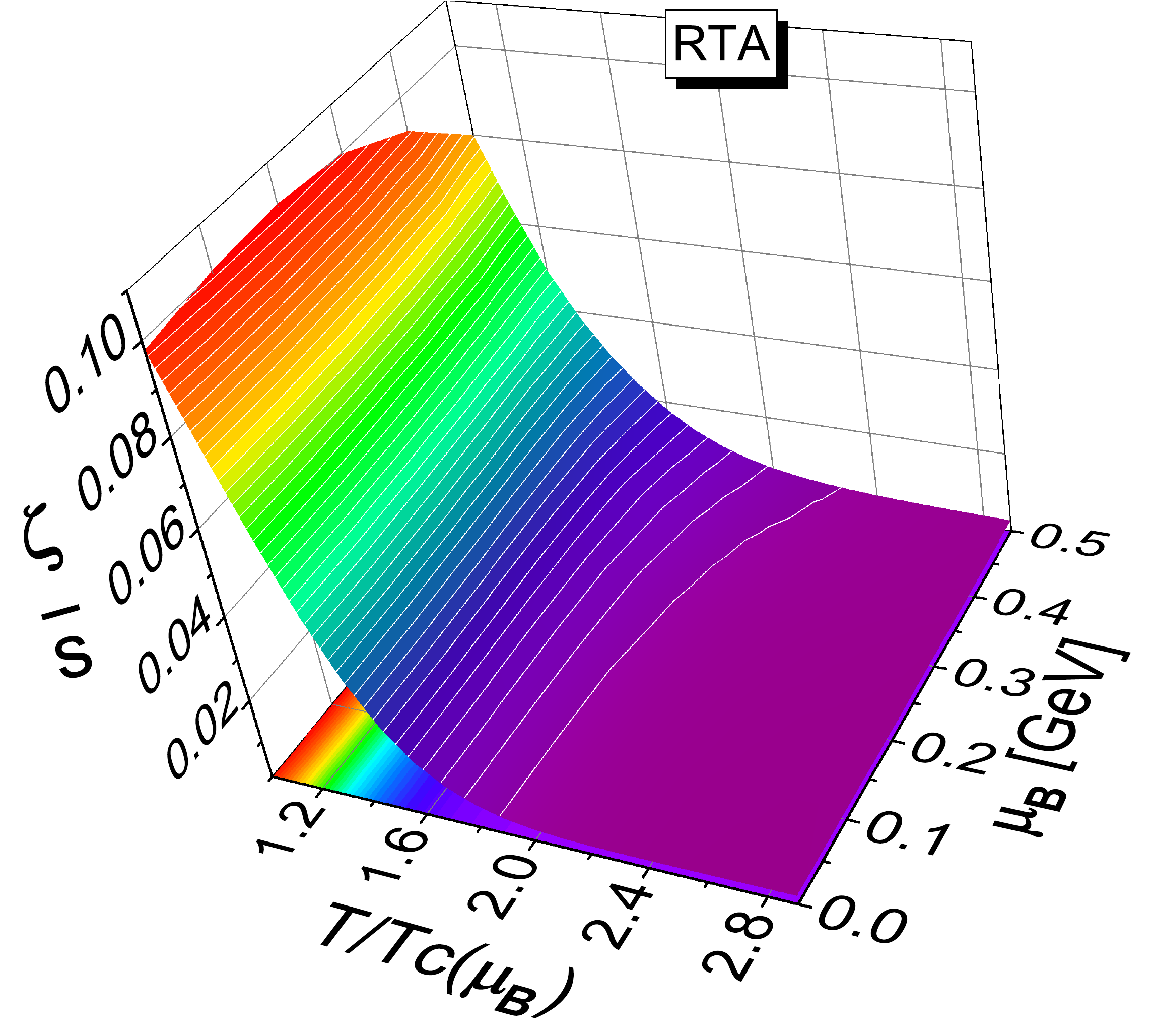} \\ b) $\mu_B$ dependence}
\end{minipage}
\caption{Ratio of  bulk  viscosity to entropy density $\zeta/s$ a) as a function of scaled temperature $T/T_c$ for $\mu_B=0$ and b) for  non-zero $\mu_B$ as a function of the scaled temperature $T/T_c(\mu_B)$ and the baryon chemical potential $\mu_B$. The solid red line and the dashed green line show the DQPM results within the RTA from Eq.~(\ref{zeta_on}) using the parton  interaction rate $\Gamma_i(\mathbf{p},T,\mu)$  and the spectral width $2\gamma_i(T,\mu)$ for the relaxation time. The symbols display the lQCD data for $N_f=0$ pure SU(3) gauge theory taken from Refs. \cite{Astrakhantsev:2018} (pentagons) and \cite{Meyer} (circles). The solid blue line shows the estimate from a Bayesian analysis of experimental heavy-ion data taken from Ref.~\cite{Xu17}.}
\label{fig:zetas}
\end{figure}

Fig.~\ref{fig:zetas} a) shows the ratio of the bulk viscosity to entropy density $\zeta/s$ as a function of  the scaled temperature $T/T_c$ for $\mu_B=0$. The solid red line $(\zeta^{\rm{RTA}}_{\Gamma^{\rm{on}}}/s)$ displays the results from Eq. (\ref{zeta_on}) using the interaction rate $\Gamma_i(\mathbf{p},T,\mu)$ while the dashed green line shows the same result in the relaxation-time approximation (\ref{zeta_on}) by replacing $\Gamma_i$ by the spectral width $2\gamma_i$. The symbols correspond to the lQCD data for pure SU(3) gauge theory taken from Refs.~\cite{Astrakhantsev:2018}(pentagons) and \cite{Meyer}(circles). The solid blue line shows the results from a Bayesian analysis of experimental heavy-ion data from Ref.~\cite{Xu17}. The bulk viscosities of our quasiparticle models are always smaller than the shear viscosities even close to the phase transition.

The DQPM results coincide with the lattice data, except the point at $T \approx T_c$ from Ref.~\cite{Astrakhantsev:2018}. This point has large error bars since lattice simulations at low temperatures require much larger statistics than simulations at higher $T$. We compare the bulk viscosity also to predictions from a Bayesian analysis of experimental data from Ref.~\cite{Xu17}. The DQPM exhibits the expected peak close to the critical temperature which is close to the Bayesian line maximum of the peak at $T_c$ with $\zeta/s \simeq 0.075$. The ratio from the Bayesian analysis shows a sudden drop to zero, which is incompatible with the small, non-zero lattice values. Furthermore, the DQPM results for viscosities are in a good agreement with the gluodynamic lattice QCD calculation at $\mu_B=0$ from Ref.~\cite{Astrakhantsev:2018}. In the case of the bulk viscosity $\zeta$ we have found that the original DQPM calculations are very close to the results obtained using the interaction rates, such that they merge in  Fig.~\ref{fig:zetas} a). The ratio $\eta/s$ increases with $\mu_B$ at all temperatures, while $\zeta/s$ only for $T > 1.2T_c$. It decreases in the vicinity of $T_c$, where the bulk viscosity is dominated by the mean-field effects that enter via $dM^2 /dT^2$. In the DQPM the masses depend primarily on the effective coupling $g^2$, which decreases as a function of $\mu_B$, also the mean-field effects become weaker. This causes a small decrease of $\zeta/s$ in contrast to the other transport coefficients. At higher temperatures the mean-field effects become also less pronounced, resulting in a decreasing $\zeta/s$ as a function of temperature. That clarifies why the $\mu_B$ behavior of the bulk viscosity changes with temperature. When the mean-field effects become subleading, their further decrease has no influence on the bulk viscosity and the ratio $\zeta/s$ starts to increase with $\mu_B$ as we will see later for the other transport coefficients.

\subsection{Electric conductivity}
Another important transport coefficient is the electric conductivity for stationary electric fields $\sigma_0$ which describes the response of the system to an external electric field. The study of the temperature and baryon chemical potential dependence of $\sigma_0$  is of fundamental importance for the possible generation of the chiral-magnetic effect in predominantly peripheral heavy-ion reactions. Moreover, $\sigma_0$  influences the emission rate of soft photons \cite{Yin}  as well as their  spectra \cite{Turbide,Akamatsu11,Linnyk13}.
The electric conductivity $\sigma_0$  is evaluated by using the relaxation time approximation (see Ref.~\cite{Thakur} for a detailed derivation):
\begin{align}
    \sigma_0^{\text{RTA}}(T,\mu_B) = \frac{e^2}{3T} \sum_{i=q,\bar{q}}  q_i^2 \int \frac{d^3p}{(2\pi)^3} \frac{\mathbf{p}^2}{E_i^2}    \label{sigm_on} \\
 \times \tau_i(\mathbf{p},T,\mu_B) d_i  (1 - f_i) f_i , \nonumber
\end{align}
where $e^2=4 \pi \alpha_{em} $, $q_i=+2/3(u),-1/3(d),-1/3(s)$ are the quark charges, $d_q = 2N_c = 6$ are degeneracy factors for spin and color in case of quarks and anti-quarks, $\tau_i$ their relaxation times, while $f_i$ denote the Fermi-Dirac distribution functions for quark and anti-quarks. In these formulae we deal with quarks and anti-quarks of $N_f=3$ flavours. Each parton has a contribution proportional to its charge squared. Unlike viscosities, the electric conductivity doesn't contain a contribution from gluons.
\begin{figure}[h]
\begin{minipage}[h]{0.85\linewidth}
\center{\includegraphics[width=0.97\linewidth]{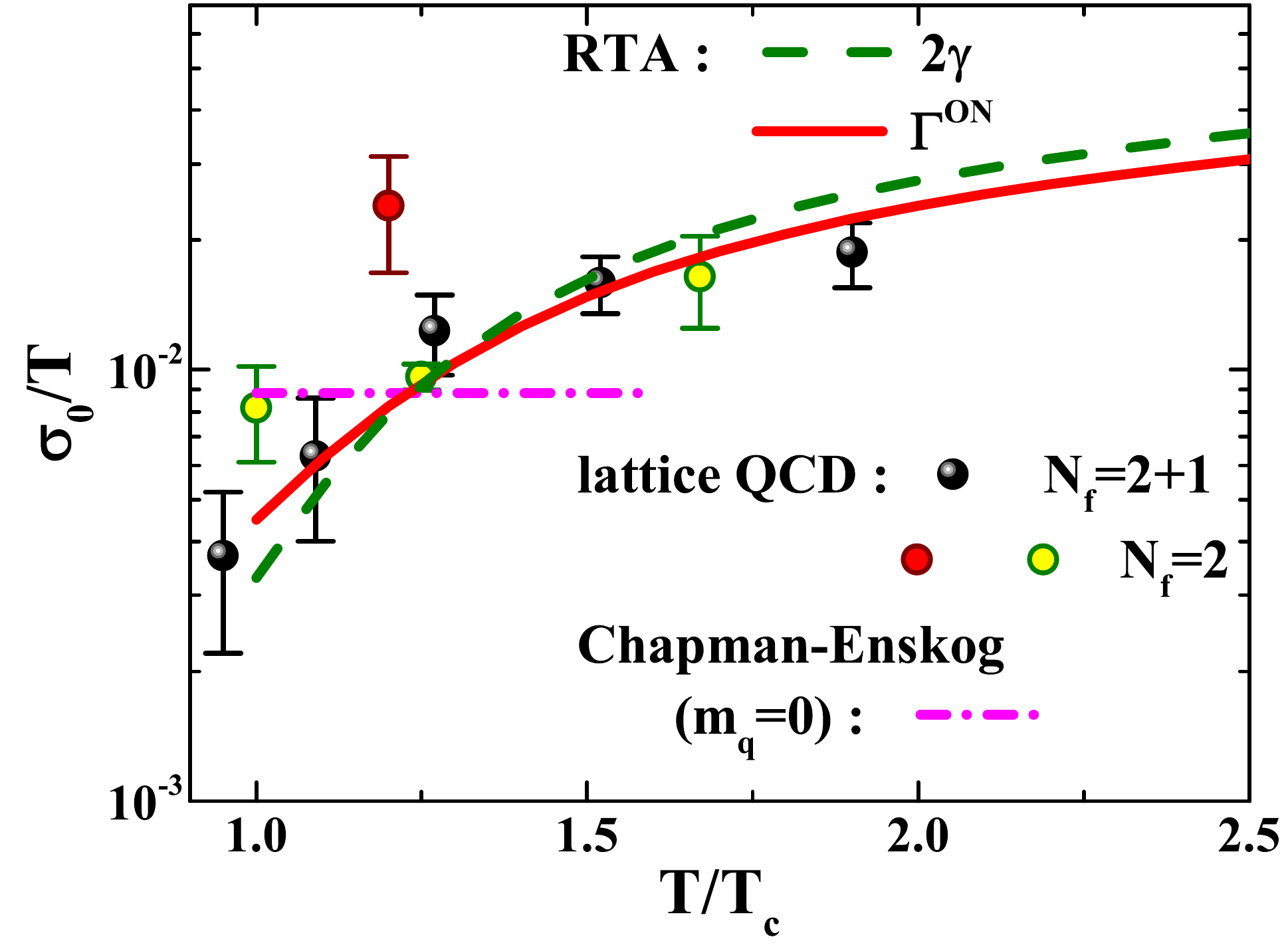} \\ a) $\mu_B$ =0 }
\end{minipage}
\hfill
\begin{minipage}[h]{0.97\linewidth}
\center{\includegraphics[width=1\linewidth]{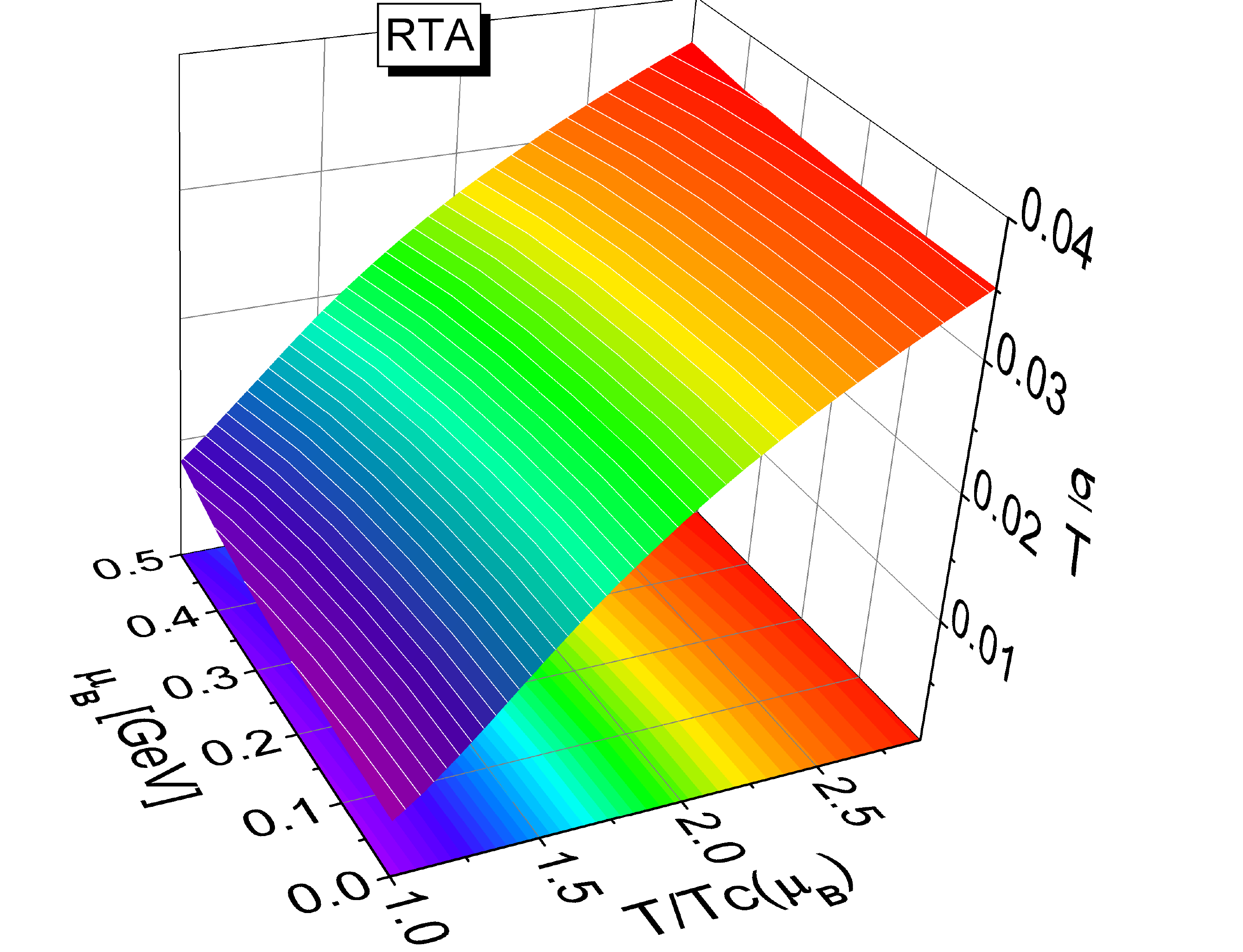} \\ b) $\mu_B$ dependence}
\end{minipage}
\caption{Ratio of electric conductivity to temperature $\sigma_0/T$ a) as a function of the scaled temperature $T/T_c$ for $\mu_B=0$ and b) for  non-zero $\mu_B$ as a function of the scaled temperature $T/T_c(\mu_B)$ and the baryon chemical potential $\mu_B$. The solid red line and the dashed green line show the DQPM results within the RTA from Eq.~\ref{sigm_on} using the parton  interaction rate $\Gamma_i(\mathbf{p},T,\mu)$  and the spectral width $2\gamma_i(T,\mu)$ for the relaxation time. The symbols display lQCD data for $N_f=2$ taken from Refs.~\cite{Brandt13,Brandt12,Brandt16} (red circles with brown borders), (yellow circles with green borders) and for $N_f=2+1$ taken from Refs.~\cite{Aarts13,Aarts15} (spheres). The dot-dashed magenta line corresponds to the results from the first order Chapman-Enskog approximation taken from Ref.~\cite{Greif18}.}
\label{fig:cond}
\end{figure}

Fig. \ref{fig:cond} depicts the  results for $\sigma_0/T$ a) as a function of the scaled temperature  and   b) the $\mu_B$ and $T$ dependence from the DQPM. The solid red line shows the DQPM result $\sigma_0^{\text{RTA}}/T$  from Eq. (\ref{sigm_on}) using the interaction rate $\Gamma_i(\mathbf{p},T,\mu)$ while the dashed green line  shows the  result in the relaxation-time approximation (\ref{sigm_on}) by replacing $\Gamma_i$ by the spectral width $2\gamma_i(T,\mu)$. We find that both values for $\sigma_0/T$ are in a good agreement. The differences between the two lines can vary by 14\%, except close to $T_c$ where the momentum dependencies of relaxation times can play a role. This approximate equivalence demonstrates again that the effective widths in the parton propagators - providing the spectral widths of the partons - are well in line with the microscopic collision rates.
The ratio rises quadratically with temperature above $T_c$ which is most likely due to the increasing number of quarks at higher temperatures. An increase in the number of charge carriers  leads to an increasing electric current and therefore to an increasing conductivity. The momentum integration smoothes the temperature dependence of the ratio.

There is another way to compute the electric conductivity by solving the relativistic transport equations for partons in a box with periodic boundary conditions in the presence of an external electric field as in Refs.~\cite{condPHSD:11,Greco14}. We compare also to the estimate from the Chapman-Enskog method using cross-sections for massless quarks and gluons as in Ref.~\cite{Greif18}, which are fixed  in order to descibe the Kovtun-Son-Starinets bound for the shear viscosity to entropy density ratio $(\eta/s)_{KSS}=1/(4 \pi)$ \cite{Kovtun:2004}, leading to $\sigma_{tot}\approx 0.72/T^2$. We find a good agreement in the vicinity of $T_c$. Furthermore, there are holographic calculations for $\sigma_0/T$ \cite{Rougemont:hologr}, which are close to our results in the vicinity of the transition $T_c-1.5T_c$, however, the  temperature dependence of the ratio differs and the values at high temperatures are lower than the DQPM predictions.

\subsection{Baryon diffusion}
It is interesting to consider further transport coefficients which are expected to be more sensitive to the net baryon density of the system, e.g.  the  baryon diffusion coefficient. This transport coefficient reveals the response to inhomogenities in the baryon density.
The baryon diffusion coefficient regulates the dissipative part of the baryon current which can be expressed as :
\begin{equation}
    \delta J^{\mu}_B= \kappa_B D^{\mu}\left(\frac{\mu_B}{T}\right),
\end{equation}
where $\kappa_B$ is the baryon diffusion coefficient, $D^{\mu}=d^{\mu}-u^{\mu}u^{\nu}d_{\nu}$ is the transverse gradient, while $u^{\mu}$ is the local fluid velocity. The dissipative baryon current can be related to the heat flow as $q^\mu= - \frac{\epsilon + p}{n_B} \delta J^{\mu}_B $ (see Refs ~\cite{Jaiswal},\cite{Son:2006}) ,
\begin{equation}
     q^\mu= \lambda \frac{n_B}{\epsilon + p} D^{\mu}\left(\frac{\mu_B}{T}\right),
\end{equation}
where $\lambda$ is the heat conductivity. Thus we can obtain a relation between the heat conductivity and the diffusion coefficient:
\begin{equation}
     \kappa_B= \lambda \left(\frac{n_BT}{\epsilon + p}\right)^2 .
     \label{heatlambda}
\end{equation}
One can easily  estimate that the values of these two coefficients differ by 2 orders of magnitude.

The diffusion coefficient can be calculated within the relaxation time approximation.
\begin{align}
\kappa_B^{\text{RTA}}(T,\mu_B) = \frac{1}{3} \sum_{i=q,\bar{q}} \int \frac{d^3p}{(2\pi)^3} \mathbf{p}^4   \tau_i(\mathbf{p},T,\mu_B) \label{kappa_RTA}\\
\times \frac{d_i  (1 \pm f_i) f_i  }{E_i^2}\left( b_a-\frac{n_B E_i}{\epsilon + p}\right)^2, \nonumber
\end{align}
where $b_i=\pm 1/3$  is the baryon number of quark and antiquark, $n_B$ is the baryon density, $w=\epsilon + p$ is the enthalpy. Taking into account relation (\ref{heatlambda}) one can see that Eq. (\ref{kappa_RTA}) is in agreement with the RTA expression for the heat conductivity as derived in Ref.~\cite{Kapusta}.

\begin{figure}[b]
\begin{minipage}[h]{0.85\linewidth}
\center{\includegraphics[width=1.\linewidth]{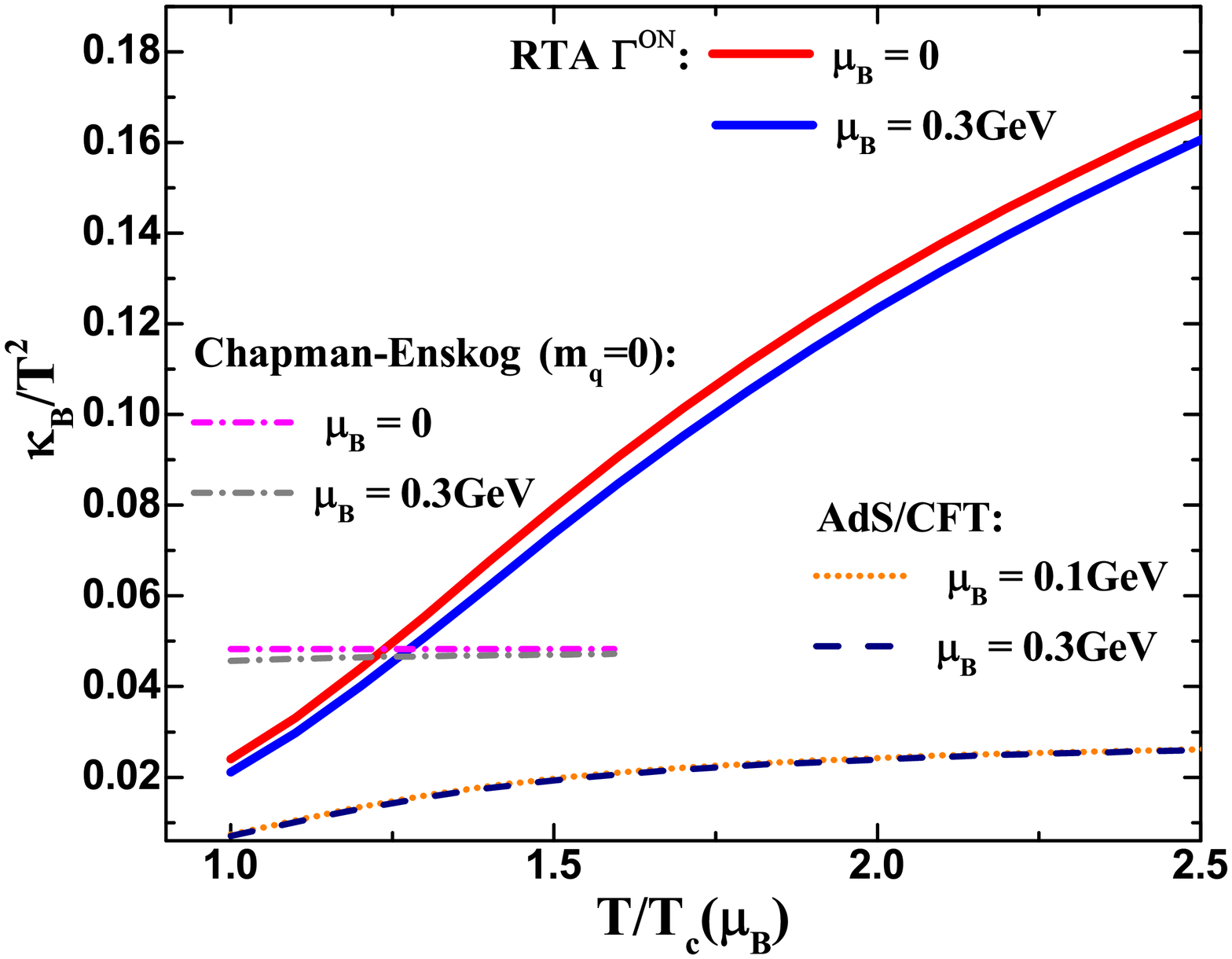} \\ a)  fixed $\mu_B$ }
\end{minipage}
\hfill
\begin{minipage}[h]{0.97\linewidth}
\center{\includegraphics[width=1.\linewidth]{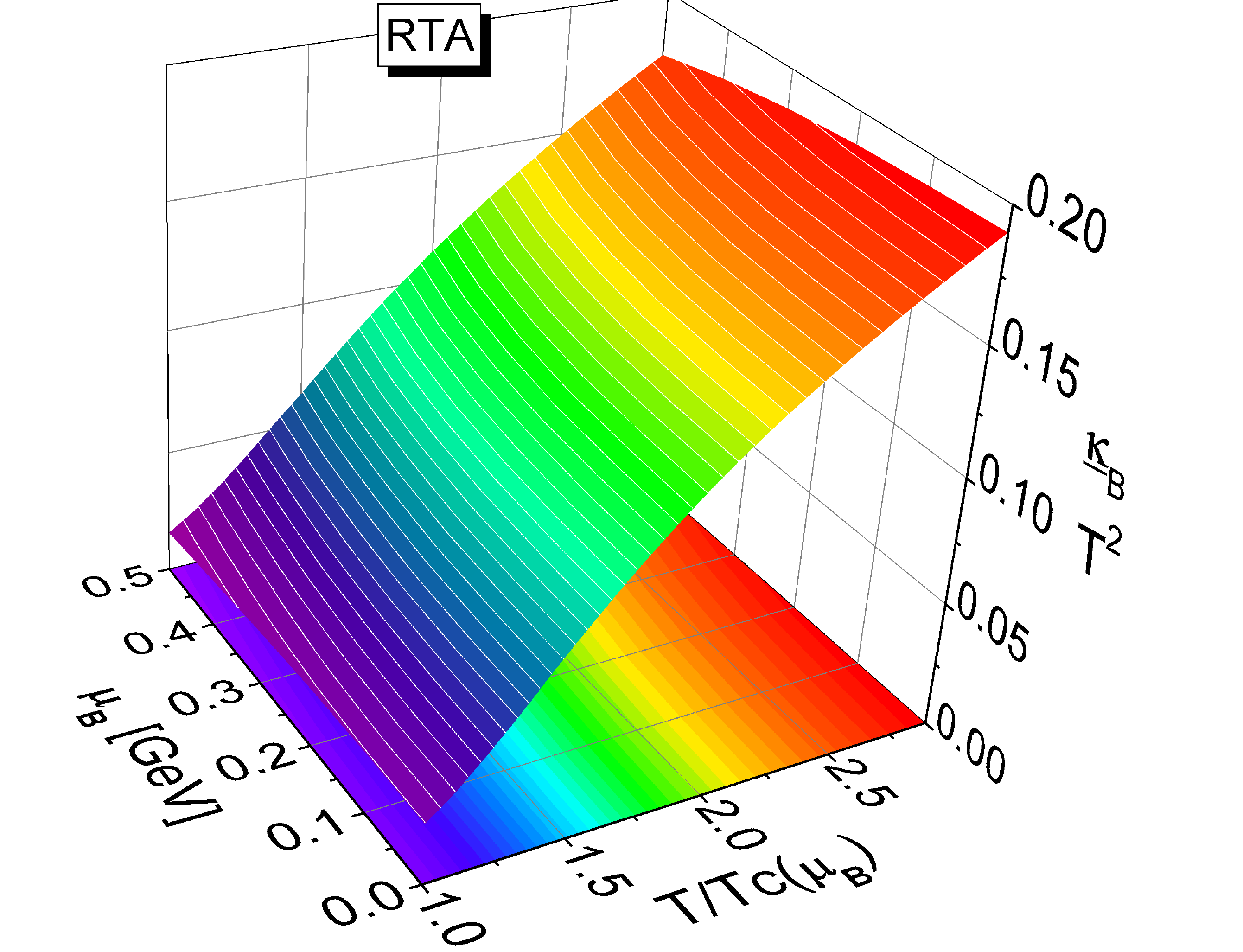} \\ b)  $\mu_B$ dependence }
\end{minipage}
\caption{Ratio of the baryon diffusion coefficent to the temperature squared $\kappa_B/T^2$ a)  as a function of scaled temperature for fixed  baryon chemical potential $\mu_B=0$ and $0.3$ GeV and b) as a function of scaled temperature for different values of the baryon chemical potential $\mu_B$. The dashed lines represent the AdS/CFT results for $\kappa_B^{SS}$, which is obtained using the results from Ref.~\cite{Son:2006} and the DQPM EoS. The dot-dashed lines correspond to the results from the first order Chapman-Enskog approximation taken from Ref.~\cite{Greif18}.}
\label{kappa_b}
\end{figure}

Fig.~\ref{kappa_b} shows the actual results for the baryon diffusion coefficent in the range of temperature and non-zero baryon chemical potential $\mu_B$. We compare the DQPM results to the estimates based on the thermal conductivity results from the AdS/CFT correspondence \cite{Son:2006}. Using the relation between the baryon diffusion coefficient and the heat conductivity (\ref{heatlambda}), we can translate these results to the following expression:
\begin{equation}
\kappa_B^{SS}= 2 \pi\frac{Ts}{\mu_B^2} \left( \frac{n_BT}{\epsilon + p}\right)^2,
\end{equation}
where $s$ is the entropy density, $n_B$ is the baryon density, $w=\epsilon + p$ is the enthalpy. We have calculated $\kappa_B^{SS}$ using  $s$, $n_B$, $\epsilon$ and $p$  from the DQPM.

In the vicinity of $T_c$ the DQPM values for the diffusion coefficient are in agreement with the calculations within the Chapman-Enskog first-order approximation using cross-sections for massless quarks and gluons in Ref.~\cite{Greif18}. However, for higher temperatures the ratio $\kappa_B^{RTA}/T^2$ grows with temperature in the DQPM while the Chapman-Enskog results stay approximately constant $\kappa_B^{CE}/T^2\sim0.048$ for all temperatures. It is expected that $\kappa_B$ has a more sizable $\mu_B$-dependence than another transport coefficients and we see a slight decrease of the ratio $\kappa_B^{RTA}/T^2$ for the DQPM with increasing chemical potential $\mu_B$, while results from other approaches  $\kappa_B^{CE}/T^2\sim0.048$  and $\kappa_B^{SS}$ are approximately $\mu_B$ independent for the considered region of $\mu_B$.
The same slight decrease has been seen in the holographic calculations in Ref.~\cite{Rougemont:hologr}.

$\kappa_B$ can be an interesting property of the partonic phase, which may have effects on  observables: it has been found within the hybryd (hydrodynamics+transport transport) theoretical framework in Ref.~\cite{Denicol} that the baryon diffusion enhances the difference between proton and antiproton mean transverse momenta and elliptic flow $v_2(p_T)$: decreases proton elliptic flow while increasing anti-proton elliptic flow. Further studies of elliptic and direct flow within the extended parton-hadron-string dynamics (PHSD) transport approach can be made to quantify the effect. In the extended version of PHSD \cite{Pierre19} the partonic sector is described by explicitly calculating the total and differential partonic scattering cross sections as a function of temperature $T$ and baryon chemical potential $\mu_B$ on the basis of the effective propagators and couplings from the DQPM and thus is a suitable extension of the DQPM to nonequilibrium configurations as encountered in relativistic heavy-ion collisions.

\section{\label{sec4}Conclusions}
This study presents the results for transport coefficients of the QGP on the basis of the dynamical quasiparticle model DQPM. We have calculated the scaled temperature $T/T_c$ and baryon chemical potential $\mu_B$ dependence of transport coefficients such as shear and bulk viscosity, electric conductivity and baryon diffusion coefficient. All calculations have been performed using the relaxation time approximation, where we have evaluated the relaxation times using a) the parton interaction rates $\tau_i(\mathbf{p},T,\mu) = \frac{1}{\Gamma_i(\mathbf{p},T,\mu)}$, where all the elastic two-body scatterings are employed, and b) the parton spectral widths  $\tau_i(T,\mu) = \frac{1}{2\gamma_i(T,\mu)}$. The results for the transport coefficients - using two different evaluations of the relaxation time - are very close to each other and differ basically only close to the critical temperature.

In case of the shear viscosity $\eta$ we showed the individual contributions of  light quarks, strange quarks and gluons for $\mu_B$=0. The total ratio of the shear viscosity to entropy density $\eta/s$ is in a good agreement with the most recent gluodynamic lQCD data and estimation from a Bayesian analysis of experimental data. This ratio grows smoothly with the baryon chemical potential for all temperatures in the considered range  $\mu_B \leq$ 0.5 GeV. 

Furthermore, we have calculated the bulk viscosity to entropy density ratio $\zeta/s$, which matches perfectly with the LQCD data at $\mu_B$ =0, while the Bayesian analysis gives smaller values. The bulk viscosity has a (slight) $\mu_B$ dependence similar to the shear viscosity except in the vicinity of $T_c$, where mean-field effects play a role.

We have considered futher transport coefficients such as the electric conductivity $\sigma_0$ and the baryon diffusion coefficient $\kappa_B$, where the gluons contribute only via the relaxation times. The dimensionless ratio of electric conductivity to temperature $\sigma_0/T$ has been found close to the lQCD results for $N_f=2+1$. In the vicinity of $T_c$ we find a good agreement between the DQPM results and the predictions from the Chapman-Enskog approximation for massless quarks and gluons. The ratio $\sigma_0/T$ shows slight increase in  $\mu_B$ similar to the viscosities.
Since there are no available lQCD calculations of $\kappa_B$, it is interesting to estimate $\kappa_B$ and its $\mu_B$-dependence. Actual DQPM results for the dimensionless ratio of the baryon diffusion coefficient to the temperature squared $\kappa_B^{RTA}/T^2$ are in agreement with the estimates from the Chapman-Enskog approximation for massless quarks and gluons near $T_c$. Moreover, we have estimated the value of  $\kappa_B^{SS}/T^2$ as suggested by the AdS/CFT approach using the DQPM equation of state, taking into account the KSS bound for the shear viscosity to entropy density ratio. The actual values for $\kappa_B^{RTA}/T^2$ are larger than $\kappa_B^{SS}/T^2$ for all considered $\mu_B$ values. 

In conclusion, we have found only a very weak dependence on $\mu_B$ for all transport coefficients considered in this study. The shear and bulk viscosities, electric conductivities of the quark-gluon plasma increase with increasing $\mu_B$ while the diffusion baryon coefficient decreases. This is interesting in light of a recent finding that the baryon diffusion might enhance the difference between proton and antiproton elliptic flow $v_2(p_T)$ and mean transverse momenta.

\begin{acknowledgments}
The authors acknowledge inspiring discussions with  W. Cassing, J. M. Torres-Rincon, J. Aichelin, H. Berrehrah, C. Ratti, E. Seifert, A. Palmese and T. Steinert.  Furthermore, we would like to thank J. A. Fotakis for sharing Chapman-Enskog results for the baryon diffusion coefficient and the electric conductivity. This work was supported  by the LOEWE center "HIC for FAIR". Furthermore, P.M., and E.B. acknowledge support by the Deutsche Forschungsgemeinschaft (DFG, German Research Foundation) through the grant CRC-TR 211 'Strong-interaction matter under extreme conditions' - Project number 315477589 - TRR 211. O.S. acknowledges support from HGS-HIRe for FAIR; E.B. thanks the COST Action THOR, CA15213.
The computational resources have been provided by the LOEWE-Center for Scientific Computing.
\end{acknowledgments}


\begin{thebibliography}{99}
\bibitem{Adamczyk:BES17}
 L. Adamczyk et al. (STAR), Phys. Rev. C {\bf 96}, 044904
(2017), arXiv:1701.07065 [nucl-ex].
\bibitem{FAIR}
 T.  Ablyazimovet et al. (CBM),  Eur.  Phys.  J. A {\bf 53},  60 (2017), arXiv:1607.01487 [nucl-ex].
\bibitem{NICA}
 A.  N.  Sissakian  and  A.  S.  Sorin (NICA), Strangeness in  quark  matter.  Proceedings,  International  Conference, SQM 2008,  Beijing,  P.R. China,  October 5-10,  2008, J. Phys. G {\bf 36}, 064069 (2009).

\bibitem{Cassing:2008nn}
W.~Cassing,
Eur.\ Phys.\ J.\ ST {\bf 168}, 3 (2009); Nucl. Phys. A {\bf 795}, 70 (2007).
\bibitem{Vanderheyden:1998}
B. Vanderheyden and G. Baym, J. Stat. Phys. {\bf 93}, 843 (1998).
\bibitem{Blaizot:2000fc}
J.~P.~Blaizot, E.~Iancu and A.~Rebhan, Phys. Rev. D {\bf 63}, 065003
(2001).
\bibitem{Sasaki:2009}
C. Sasaki and K. Redlich, Phys. Rev. C {\bf 79}, 055207 (2009).
\bibitem{Bluhm:2011}
M. Bluhm, B. K\"{a}mpfer, and K. Redlich, Phys. Rev. C {\bf 84}, 025201 (2011).
\bibitem{Marty:NJL13}
R. Marty, E. Bratkovskaya, W. Cassing , J. Aichelin , H. Berrehrah,
Phys. Rev. C {\bf 88}, 045204 (2013).
\bibitem{Pierre19}
P. Moreau, O. Soloveva, L. Oliva, T. Song, W. Cassing, and E. Bratkovskaya, Phys. Rev. C {\bf 100}, 014911 (2019).
\bibitem{Mykhaylova:2019}
V. Mykhaylova, M. Bluhm, K. Redlich, and C. Sasaki, Phys. Rev. D {\bf 100}, 034002 (2019).
\bibitem{Review}
O. Linnyk, E. L. Bratkovskaya, and W. Cassing, Prog. Part. Nucl. Phys. {\bf 87}, 50 (2016).
\bibitem{Hamsa:PRC16}
H. Berrehrah, W. Cassing, E. Bratkovskaya, and T. Steinert,
Phys. Rev. C {\bf 93}, 044914 (2016).
\bibitem{Hamsa:JModPhys16}
H. Berrehrah, E. Bratkovskaya, T. Steinert, and W. Cassing,
Int. J. Mod. Phys. E {\bf 25}, 1642003 (2016).

\bibitem{Borsanyi:2013bia}
S.~Borsanyi, Z.~Fodor, C.~Hoelbling, S.~D.~Katz, S.~Krieg and K.~K.~Szabo,
Phys.\ Lett.\ B {\bf 730}, 99 (2014).

\bibitem{Borsanyi:2012cr}
S.~Borsanyi, G.~Endr\"{o}di, Z.~Fodor, S.~D.~Katz, S.~Krieg, C.~Ratti and K.~K.~Szabo,
JHEP {\bf 1208}, 053 (2012).




\bibitem{Thorsten}  	
T. Steinert and W. Cassing,  J. Phys. Conf. Ser. 1024, 012029 (2018).
\bibitem{Cleymans:2006}
J. Cleymans, H. Oeschler, K. Redlich, and S. Wheaton, Phys. Rev. C {\bf 73}, 034905 (2006).
\bibitem{Palmese:2016}
A. Palmese, W. Cassing, E. Seifert, T. Steinert, P. Moreau, and E. L. Bratkovskaya, Phys. Rev. C {\bf 94}, 044912 (2016).

\bibitem{Hosoya:RTA}
A. Hosoya and K. Kajantie, Nucl. Phys. B {\bf 250}, 666 (1985).
\bibitem{Kubo}
R. Kubo, J. Phys. Soc. Jpn.  {\bf 12}, 570 (1957).
\bibitem{Aarts:2002}
G. Aarts and J. M. Martinez Resco, J. High Energy Phys. {\bf 04}, 053 (2002).
\bibitem{Chakraborty11}
P. Chakraborty and J. I. Kapusta, Phys. Rev. C {\bf 83}, 014906 (2011).
\bibitem{Kapusta}
M. Albright and J. I. Kapusta, Phys. Rev. C {\bf 93}, 014903 (2016).

\bibitem{SGavin}
S. Gavin, Nucl. Phys. A {\bf 435}, 826 (1985).
\bibitem{Fraile:2006}
D. F. Fraile, A. G. Nicola, Phys. Rev. D {\bf 73}, 045025 (2006).
\bibitem{Lang12}
R. Lang, N. Kaiser, and W. Weise, Eur. Phys. J. A {\bf 48}, 109 (2012).
\bibitem{Ozvenchuk13:kubo}
V. Ozvenchuk, O. Linnyk, M. I. Gorenstein, E. L. Bratkovskaya, and W. Cassing, Phys. Rev. C {\bf 87}, 064903 (2013).
\bibitem{Shuryak} 
E. Shuryak, Prog. Part. Nucl. Phys. {\bf 53}, 273 (2004).
\bibitem{Gyulassy}
M. Gyulassy and L. McLerran, Nucl. Phys. A 750, 30 (2005).
\bibitem{Heinz}
U. W. Heinz, arXiv:nucl-th/0512051.
\bibitem{Astrakhantsev:2017}
N. Astrakhantsev, V. Braguta, and A. Kotov, J. High Energy Phys. {\bf 04}, 101 (2017).
\bibitem{Sbass:2017}
S. A. Bass, J. E. Bernhard and J. S. Moreland, Nucl. Phys. A {\bf 967}, 67 (2017)
\bibitem{Ryu}
S. Ryu, J. - F. Paquet, C. Shen, G. S. Denicol, B. Schenke, S. Jeon, and C. Gale, Phys. Rev. Lett. {\bf 115}, 132301 (2015).
\bibitem{Astrakhantsev:2018}
N Yu. Astrakhantsev, V.V. Braguta, and  A. Yu. Kotov, Phys.Rev. D {\bf 98} no.5, 054515 (2018).
\bibitem{Meyer}
H. B. Meyer, Phys. Rev. Lett. {\bf 100}, 162001 (2008).
\bibitem{Xu17}
Y. Xu, P. Moreau, T. Song, M. Nahrgang, S. A. Bass, and E. Bratkovskaya, Phys. Rev. C {\bf 96}, no. 2, 024902 (2017).

\bibitem{Steinert:14}
T. Steinert and W. Cassing, Phys. Rev. C {\bf 89}, no. 3, 035203 (2014).
\bibitem{condPHSD:11}
W. Cassing, O. Linnyk, T. Steinert, and V. Ozvenchuk, Phys. Rev. Lett. {\bf 110}, no.
18, 182301 (2013).
\bibitem{Yin}
Y. Yin, Phys. Rev. C {\bf 90}, 044903 (2014).
\bibitem{Turbide}
S. Turbide, R. Rapp, and C. Gale, Phys. Rev. C {\bf 69}, 014903 (2004).
\bibitem{Linnyk13}
O. Linnyk, W. Cassing, and E. Bratkovskaya, Phys. Rev. C {\bf 89},  (2013).
\bibitem{Akamatsu11}
Y. Akamatsu, H. Hamagaki, T. Hatsuda, and T. Hirano, J. Phys. G {\bf 38}, 124184 (2011).
\bibitem{Greco14}
A. Puglisi, S. Plumari, and V. Greco, Phys. Rev. D {\bf 90}, 114009 (2014).
\bibitem{Thakur}
L. Thakur, P. K. Srivastava, G. P. Kadam, M. George, and H. Mishra,  Phys. Rev. D {\bf 95}, 096009 (2017).
\bibitem{Brandt13}
B. B. Brandt, A. Francis, H. B. Meyer, and H. Wittig, JHEP {\bf 1303}, 100 (2013).
\bibitem{Brandt12}
B. B. Brandt, A. Francis, H. B. Meyer, and H. Wittig, PoS ConfinementX, 186 (2012).
\bibitem{Brandt16}
B. B. Brandt, A. Francis, B. J\"{a}ger, and H. B. Meyer, Phys. Rev. D {\bf 93}, no. 5,
054510 (2016).
\bibitem{Aarts13}
A. Amato, G. Aarts, C. Allton, P. Giudice, S. Hands, and J. I. Skullerud, Phys.
Rev. Lett. {\bf 111}, no. 17, 172001 (2013).
\bibitem{Aarts15}
 G. Aarts, C. Allton, A. Amato, P. Giudice, S. Hands, and J. I. Skullerud, JHEP
{\bf 1502}, 186 (2015).
\bibitem{Greif18}
 M. Greif, J. A. Fotakis, G. S. Denicol, and C. Greiner, Phys. Rev. Lett. {\bf 120}, 242301 (2018).
\bibitem{Kovtun:2004}
P. K. Kovtun, D. T. Son, and A. O. Starinets, Phys. Rev. Lett. {\bf 94}, 111601 (2005).
\bibitem{Rougemont:hologr}
R. Rougemont, R. Critelli, J. Noronha-Hostler, J. Noronha, and C. Ratti, Phys.Rev. D {\bf 96} no.1, 014032 (2017).

\bibitem{Moreau}
P. Moreau, O. Linnyk, W. Cassing, and E. Bratkovskaya, Phys. Rev. C {\bf 93}, 044916 (2016).
\bibitem{Ozvenchuk13}
 V. Ozvenchuk, O. Linnyk, M. I. Gorenstein, E. L. Bratkovskaya, and W. Cassing, Phys. Rev. C {\bf 87}, 024901 (2013).



\bibitem{Son:2006}
D. T. Son and A. O. Starinets, J. High Energy Phys. {\bf 0603}, 052 (2006).
\bibitem{Jaiswal}
A. Jaiswal, B. Friman, and K. Redlich , Phys. Lett. B {\bf 751}, 548-552 (2015).
\bibitem{Denicol}
G. S. Denicol, C. Gale, S. Jeon, A. Monnai, B. Schenke, and C. Shen, Phys. Rev. C {\bf 98}, 034916
(2018).


\end{thebibliography}
\end{document}